\documentclass{ws-jai}
\usepackage[flushleft]{threeparttable}

\usepackage[utf8]{inputenc}
\usepackage[T1]{fontenc}
\usepackage[british]{babel}
\usepackage{times}
\usepackage{setspace}

\usepackage{upgreek}
\usepackage{amsfonts,amsmath,amssymb}
\usepackage{empheq}
\usepackage{commath}
\usepackage{aas_macros}

\usepackage{ragged2e}
\usepackage{multicol}
\usepackage{framed}
\usepackage{float}
\usepackage{color}   
\usepackage{url}
\usepackage{etoolbox}
\usepackage{tabularx}
\usepackage{caption}
\usepackage{subcaption}

\usepackage{hyperref}
\hypersetup{
    pdffitwindow=true,
    pdfstartview={FitH},
    pdftitle={},
    pdfauthor={Pranav Sanghavi},
    pdfcreator={Pranav Sanghavi},
    pdfproducer={Pranav Sanghavi},
    pdfkeywords={
        Astronomy, Astrophysics, VLBI, Interferometry, Telescope, Radio
        },
    pdfnewwindow=true,
    colorlinks=true,
    linkcolor=blue,
    linktoc=all,
    citecolor=red,
    urlcolor=magenta
    }

\usepackage{graphicx} 
\graphicspath{{figures/}} 

\newcommand{\seefig}[1]{(see Figure~\ref{#1})}

\newcommand{\kb}[0]{k_\mathrm{B}}

\newcommand{\Tsys}[0]{T_\mathrm{sys}}

\newcommand{\fwhm}[0]{\theta_{\mathrm{FWHM}}}

\usepackage[binary-units]{siunitx}[=v2]
\DeclareSIUnit{\year}{yr}
\DeclareSIUnit{\arcmin}{arcmin}
\DeclareSIUnit{\parsec}{pc}
\DeclareSIUnit{\erg}{erg}
\DeclareSIUnit{\mas}{mas} 
\DeclareSIUnit{\arcsec}{asec}
\DeclareSIUnit{\jansky}{Jy}
\DeclareSIUnit{\FRBs}{FRBs}
\DeclareSIUnit{\FRB}{FRB}
\DeclareSIUnit{\sky}{sky}
\DeclareSIUnit{\day}{day}
\DeclareSIUnit{\month}{month}
\DeclareSIUnit{\ethernet}{E}
\DeclareSIUnit{\deg}{deg}
\DeclareSIUnit{\TECu}{TECu}
\DeclareSIUnit{\electron}{electron}
\DeclareSIUnit{\MSPS}{Msps}
\usepackage{tikz}
\usetikzlibrary{calc}
\usetikzlibrary{patterns}
\usetikzlibrary{shapes.geometric}
\usetikzlibrary{shapes}
\usepackage{tikz-3dplot}
\usepackage{tkz-euclide}

\pgfdeclaredecoration{fiber}{initial}
{
    \state{initial}[width=1.5\pgfdecorationsegmentamplitude, next state=final]{
    \pgfpathmoveto{\pgfpointorigin}
    \pgflineto{\pgfpoint{\pgfdecoratedinputsegmentlength/2}{0pt}}
    \pgfpathcircle{\pgfpoint{\pgfdecoratedinputsegmentlength/2}{.5\pgfdecorationsegmentamplitude}}{.5\pgfdecorationsegmentamplitude}
    \pgfpathcircle{\pgfpoint{\pgfdecoratedinputsegmentlength/2-.25\pgfdecorationsegmentamplitude}{.5\pgfdecorationsegmentamplitude}}{.5\pgfdecorationsegmentamplitude}
    \pgfpathcircle{\pgfpoint{\pgfdecoratedinputsegmentlength/2+.25\pgfdecorationsegmentamplitude}{.5\pgfdecorationsegmentamplitude}}{.5\pgfdecorationsegmentamplitude}
    \pgfpathmoveto{\pgfpoint{\pgfdecoratedinputsegmentlength/2}{0pt}}
    }
    \state{final}{
        \pgfpathlineto{\pgfpointdecoratedpathlast}
    }
}
\begin{document}

\catchline{}{}{}{}{} 

\markboth{Pranav Sanghavi}{TONE: A CHIME/FRB Outrigger Pathfinder for localizations of Fast Radio Bursts using Very Long Baseline Interferometry}

\title{TONE: A CHIME/FRB Outrigger Pathfinder for localizations of Fast Radio Bursts using Very Long Baseline Interferometry}

\author{
Pranav Sanghavi$^{1,2,3}$,
Calvin Leung$^{4,5,6}$,
Kevin Bandura$^{2,3}$,
Tomas Cassanelli$^{7}$,
Jane Kaczmarek$^{8,9}$,
Victoria M.~Kaspi$^{10,11}$,
Kholoud Khairy$^{2,3}$,
Adam Lanman$^{10,11}$,
Mattias Lazda$^{10}$,
Kiyoshi W.~Masui$^{4,5}$,
Juan Mena-Parra$^{12,13}$,
Daniele Michilli$^{4,5}$,
Ue-Li Pen$^{14,15,16,12,17}$,
Jeffrey B.~Peterson$^{18}$,
Mubdi Rahman$^{19}$,
Vishwangi Shah$^{10,11}$
}
\address{
$^{1}$Department of Physics, Yale University, New Haven, CT 06520, USA\\
$^{2}$Lane Department of Computer Science and Electrical Engineering, 1220 Evansdale Drive, PO Box 6109 Morgantown, WV 26506, USA\\
$^{3}$Center for Gravitational Waves and Cosmology, West Virginia University, Chestnut Ridge Research Building, Morgantown, WV 26505, USA\\
$^{4}$MIT Kavli Institute for Astrophysics and Space Research, Massachusetts Institute of Technology, 77 Massachusetts Ave, Cambridge, MA 02139, USA\\
$^{5}$Department of Physics, Massachusetts Institute of Technology, 77 Massachusetts Ave, Cambridge, MA 02139, USA\\
$^{6}$NHFP Einstein Fellow\\
$^{7}$Department of Electrical Engineering, Universidad de Chile, Av. Tupper 2007, Santiago 8370451, Chile\\
$^{8}$Dominion Radio Astrophysical Observatory, Herzberg Research Centre for Astronomy and Astrophysics, National Research Council Canada, PO Box 248, Penticton, BC V2A 6J9, Canada\\
$^{9}$CSIRO Space and Astronomy, P.O. Box 76, Epping, NSW 1710, Australia\\
$^{10}$Department of Physics, McGill University, 3600 rue University, Montr\'eal, QC H3A 2T8, Canada\\
$^{11}$Trottier Space Institute, McGill University, 3550 rue University, Montr\'eal, QC H3A 2A7, Canada\\
$^{12}$Dunlap Institute for Astronomy \& Astrophysics, University of Toronto, 50 St.~George Street, Toronto, ON M5S 3H4, Canada\\
$^{13}$David A.~Dunlap Department of Astronomy \& Astrophysics, University of Toronto, 50 St.~George Street, Toronto, ON M5S 3H4, Canada\\
$^{14}$Institute of Astronomy and Astrophysics, Academia Sinica, Astronomy-Mathematics Building, No. 1, Sec. 4, Roosevelt Road, Taipei 10617, Taiwan\\
$^{15}$Canadian Institute for Theoretical Astrophysics, 60 St.~George Street, Toronto, ON M5S 3H8, Canada\\
$^{16}$Canadian Institute for Advanced Research, MaRS Centre, West Tower, 661 University Avenue, Suite 505 \\
$^{17}$Perimeter Institute for Theoretical Physics, 31 Caroline Street N, Waterloo, ON N25 2YL, Canada\\
$^{18}$Department of Physics, Carnegie Mellon University, 5000 Forbes Avenue, Pittsburgh, 15213, PA, USA\\
$^{19}$Sidrat Research, 124 Merton Street, Suite 507, Toronto, ON M4S 2Z2, Canada\\
}

\maketitle

\corres{$^{1}$Corresponding author.}

\begin{history}
\received{(to be inserted by publisher)};
\revised{(to be inserted by publisher)};
\accepted{(to be inserted by publisher)};
\end{history}

\begin{abstract}
The sensitivity and field of view of the Canadian Hydrogen Intensity Mapping Experiment (CHIME) has enabled its fast radio burst (FRB) backend to detect thousands of FRBs. However, the low angular resolution of CHIME prevents it from localizing most FRBs to their host galaxies. Very long baseline interferometry (VLBI) can readily provide the subarcsecond resolution needed to localize many FRBs to their hosts. Thus we developed TONE: an interferometric array of eight 6-m dishes to serve as a pathfinder for the CHIME/FRB Outriggers project, which will use wide field-of-view cylinders to determine the sky positions for a large sample of FRBs, revealing their positions within their host galaxies to subarcsecond precision. In the meantime, TONE's \SI{\sim3333}{\kilo\meter} baseline with CHIME proves to be an excellent testbed for the development and characterization of single-pulse VLBI techniques at the time of discovery. This work describes the TONE instrument, its sensitivity, and its astrometric precision in single-pulse VLBI. We believe that our astrometric errors are dominated by uncertainties in the clock measurements which build up between successive Crab pulsar calibrations which happen every $\approx \SI{24}{\hour}$; the wider fields-of-view and higher sensitivity of the Outriggers will provide opportunities for higher-cadence calibration. At present, CHIME-TONE localizations of the Crab pulsar yield systematic localization errors of \SIrange{0.1}{0.2}{\arcsec} -- comparable to the resolution afforded by state-of-the-art optical instruments ($\sim$\SI{0.05}{\arcsec}). 
\end{abstract}


\section{Introduction}

Fast radio bursts (FRBs) are bright, millisecond-duration transients detected at radio wavelengths~\citep[e.g.][]{Lorimer}. Their integrated column density of free electrons along the line-of-sight, known as the dispersion measures (DMs) suggest an extragalactic origin, and their characteristic radio luminosity is orders of magnitude larger than those of other known radio transients. 

Since the first FRB was reported, hundreds of FRBs have been discovered by instruments across the globe\footnote{\url{https://www.herta-experiment.org/frbstats/}}~\citep[see][]{review2022}. The vast majority of detections have been made by the Canadian Hydrogen Intensity Mapping Experiment Fast Radio Burst (CHIME/FRB) project~\cite{chimefrbcatalog}. CHIME is a novel radio telescope operating at the Dominion Radio Astrophysical Observatory (DRAO) in Penticton, British Columbia, Canada. It consists of four $\SI{20}{\m} \times \SI{100}{\m}$ fixed cylindrical reflectors oriented in the north/south direction, each equipped with \num{256} dual-polarization feeds sensitive in the range \SIrange{400}{800}{\mega\hertz}. The cylinders form a transit interferometer with \SI{\sim200}{\deg\squared} field-of-view that continuously surveys the northern half of the sky \footnote{The CHIME primary beam covers declinations down to \SI{-11}{\deg}}. The CHIME/FRB experiment is a specialized backend that triggers on detection of high-dispersion radio transients to search for FRBs in real time~\cite{chimefrbbeamfoming, chimesystem}. CHIME/FRB is capable of localizing FRBs with \SI{\sim1}{\arcminute} precision using buffered raw voltage data that can be saved for events detected by the real--time pipeline with a threshold signal to noise ratio (S/N), currently set to 12~\cite{chimebaseband}.

A small ($\sim3$\%) fraction of detected FRBs are seen to repeat \cite{repeaternature2, RN1, RN2, RN3}. At least two repeaters show long-term periodic activity~\cite{periodicrepeater, periodicity121102_1,periodicity121102_2}. Follow-up studies of FRBs are limited and very challenging due to the difficulty of associating FRBs with their host galaxies. While repeating sources can in principle eventually be localized using traditional interferometric techniques~\citep[e.g.][]{repeaterloc1}, precise sky localization has been challenging for the majority of one-off sources detected to date. This is because CHIME/FRB has an insufficient angular resolution to permit unambiguous host galaxy identification for the vast majority of its detections. Although some hosts can be identified for very bright FRBs with very low DM excess by imposing a prior on the host galaxy’s maximum redshift~\cite{localfrb2,localfrb1, hostgalaxiesofrepeaters}, too few host galaxies have been identified to discern trends in host type or of source location within the host \cite{askaphost, dsahost1, dsahost2, hostsonoutskirts, highreshosts, characterizinghost, hostgalaxyofsomerepeater, anfrbinstarforming, frbstarfomringmaybe, frboffset, frbhostvsstellartransients, spiralfrb, nimmo2022milliarcsecond}.

Very long baseline interferometry (VLBI) is a natural way to expand upon CHIME/FRB's localization capabilities. However, VLBI observations of one-off FRBs are a challenging endeavor. The broad-band emission from FRBs is, despite being confined to milliseconds, dispersed in time typically over several seconds due to its propagation through cold plasma. Their extremely long dispersive sweeps and unpredictable occurrence in time and sky location make it challenging to localize these bursts with traditional VLBI techniques of data acquisition and calibration. Traditional approaches to VLBI for FRBs have been implemented albeit only for repeaters~\cite{evnfrb, nimmo2022milliarcsecond,repeaterglobular}. A few apparently non-repeating FRBs, or `one offs', have been localized to their hosts using connected-element interferometers such as the Australian Square Kilometer Array Pathfinder (ASKAP)~\cite{ascapfrb}, Deep Synoptic Array-10 (DSA-10)~\cite{dsa10frb}, Deep Synoptic Array-110 (DSA-110)~\cite{2023AAS...24123901R}, MeerKAT~\cite{meerkat} and the Very Large Array (VLA)~\cite{vlafrb}. Nevertheless, obtaining sub-arcsecond localizations of one-off FRBs on a routine basis for a large number of sources is still a major goal of the field which remains elusive. The CHIME/FRB Outriggers project plans to fill in the gap of precise localizations of one-off FRBs.

TONE, as an interferometric array, is one of three pathfinders for the CHIME/FRB Outriggers project whose aim is to demonstrate key hardware and software instrumentation with the goal of localizing FRBs using VLBI. The CHIME Pathfinder telescope~\cite{pathfinderpaper} -- the precursor to CHIME -- was re-commissioned as a high-bandwidth data recorder and in-beam calibration strategies, which was tested on a short baseline~\cite{pathfinderoutrigger}. Additionally, we also commissioned a single-dish outrigger at the Algonquin Radio Observatory (ARO) using the 10-meter dish (ARO10)~\cite{2022AJ....163...65C} as a simple long-baseline outrigger. TONE combines the instrumentation challenges of widefield, ``large-N, low-D''  interferometric telescopes with the calibration challenges of VLBI on long baselines.

This paper is organized as follows. We describe the TONE array at Green Bank Observatory in \S\ref{sec:tone}, its analog chain in \S\ref{sec:analog}, and the digital system in \S\ref{sec:digital}.  The performance of the array is discussed in \S\ref{sec:performance}. The operations of the array are described in \S\ref{sec:ops} leading to the first light and finally beamforming. VLBI cross-correlation and localization are described in \S\ref{subsect:xcorr}. We use Crab pulsar (PSR B0531+21) giant pulse data and VLBI to empirically determine our systematic uncertainties in \S\ref{sect:systematicerror} before concluding in \S\ref{sect:dnc}.

\section{TONE}\label{sec:tone}

\begin{figure}[htp]
     \centering
     \includegraphics[width=\linewidth]{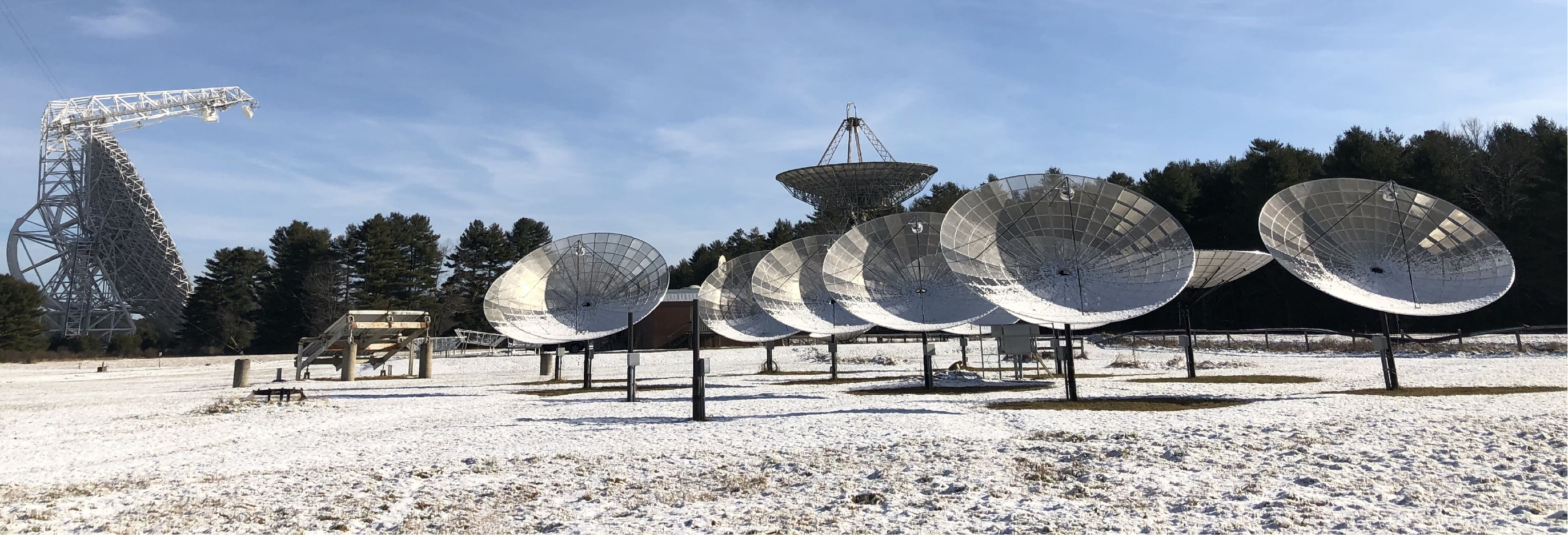}
     \caption[TONE]{TONE: The array of dishes at the Green Bank Observatory with the Green Bank Telescope and the second 85--ft Green Bank Interferometer dish in the background. This picture was taken on December 15, 2020.}
    \label{fig:tonepics}
\end{figure}

TONE (see Figure~\ref{fig:tonepics}) was built with the primary goal of detecting pulses from FRBs and localizing one-off pulses to sub-arcsecond accuracy upon detection. A secondary goal was to demonstrate key aspects of the hardware and software required for CHIME/FRB Outriggers.  Additionally, it has served as a testbed for technologies that may enable future ``large-N, low-D'' instruments such as the Hydrogen Intensity and Real-time Analysis eXperiment (HIRAX)~\cite{hirax1,hirax2} and the Canadian Hydrogen Observatory and Radio-transient Detector (CHORD)~\cite{chord}. 

TONE is located at the Green Bank Observatory in West Virginia near the control building of the Green Bank interferometer. The National Radio Quiet Zone provides a clean environment free of radio frequency interference (RFI). Existing radio astronomical infrastructure allows for resources such as buildings to house the backend, a gigabit fiber internet link, and a copy of the Green Bank maser signal. The key instrumental parameters of the array are described in Table \ref{tab:toneparameters}.

\begin{figure}[htp]
    \centering
    \includegraphics[width=0.5\textwidth]{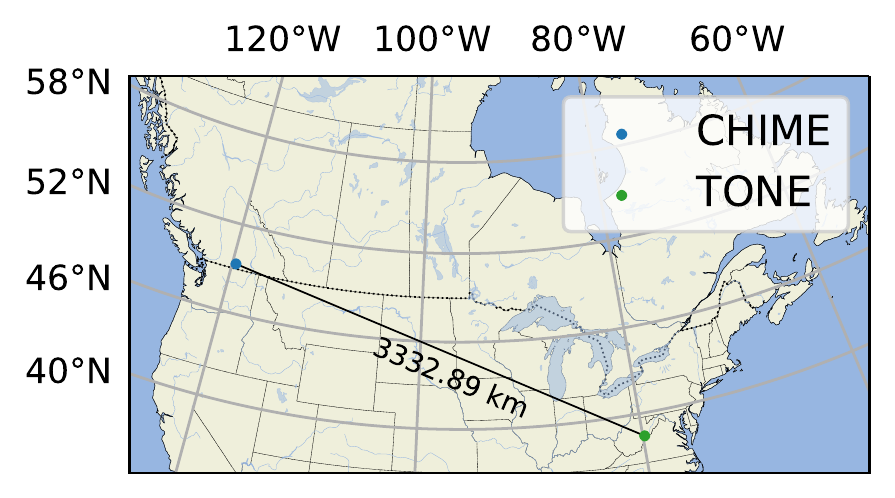}
    \caption[CHIME and TONE on a map]{Location of the CHIME and TONE. The baseline distance between CHIME and TONE is \SI{\approx3333}{\kilo\meter}.  }
    \label{fig:map}
\end{figure}

CHIME-TONE VLBI works by operating each VLBI station as a phased array and cross-correlating synthesized voltage beams from each station. The diffraction limited angular resolution over the $B_\mathrm{km} \sim \SI{3333}{\kilo\meter}$ baseline between CHIME and TONE (see Figure~\ref{fig:map}) is $\fwhm \mathrm{(in~milliarcseconds)} \sim 2063 \times {\lambda_\mathrm{cm}}/{B_\mathrm{km}},$ which at $\SI{400}{\mega\hertz} \equiv \lambda_\mathrm{cm} = \SI{75}{\centi\meter}$ gives $\fwhm \sim \SI{50}{\mas}$. This precision is sufficient to localize the majority of FRBs within their host galaxies for detailed follow-up of their environments~\cite{eftekhari2017associating}. 

\begin{table}[htb]
\centering
\begin{tabular}{l|l}
\hline
\textbf{Parameter}              & \textbf{Value}                            \\ \hline
Number of Dishes                & \num{8}         \\
Frequency Range                 & \SIrange{400}{800}{\mega\hertz}           \\ 
Frequency Resolution            & \SI{390}{\kilo\hertz}, \num{1024} channels\\ 
Dish Diameter                   & \SI{6}{\m}                                \\ 
Dish focal ratio ($f/D$)         & $\sim0.4$                                \\ 
Planned  Layout &  $4\times3$ with \SI{9.1}{\m} spacing \\ 
Primary Beam FWHM               & \SIrange{11}{5}{\degree} across the \SIrange{400}{800}{\mega\hertz} frequency band \\ \hline
\end{tabular}
\caption[TONE parameters]{Instrumental parameters for TONE.}\label{tab:toneparameters}
\end{table}

\begin{figure}[htp]
     \centering
     
    \begin{subfigure}[t]{0.5\textwidth}
         \centering
         \includegraphics[width=0.85\textwidth]{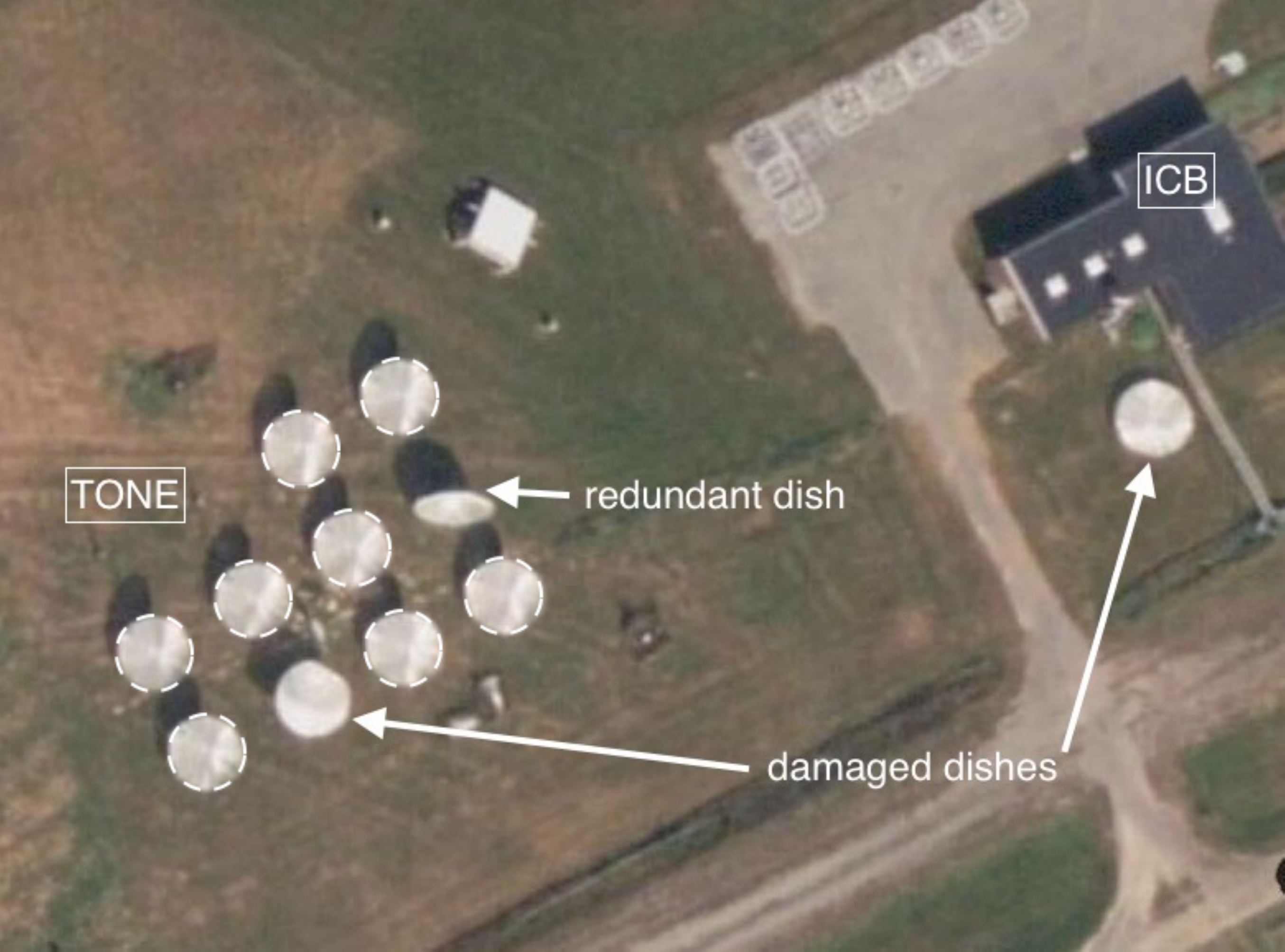}
         \caption{Satellite imagery of the TONE dishes Credit: Bing Maps}
     \end{subfigure}
     \hfill
     \begin{subfigure}[t]{0.495\textwidth}
     \centering
     \resizebox{0.75\textwidth}{!}
    	{    
    	\begin{tikzpicture}[scale=0.3]
	\edef\D{6}
	\edef\dish{
	(0.0, 0.0),
	(-8.4, 16.14),
	(-8.07, -4.2), 
	(-12.27, 3.87),
    (-16.48, 11.94),
    (-20.35, -0.33),
    (-24.22, -12.61),
    (-28.42, -4.53)}
	\coordinate (O) at (0, 0);  

	\foreach \p in \dish {
    \filldraw[color=red!60, fill=red!5, ultra thick]\p circle (3);
    }
    
    \draw[thick, loosely dotted](-15.9, -8.48) circle (3);
    \draw[thick, dotted](-4, 7.9) circle (3);

    \draw[|<->|] (-8.4-3, 16.14) -- (-8.4+3, 16.14) node[midway, fill=red!5] {$6\rm{~m}$};
    
    \draw[|<->|] (0,0) -- (-8.07, -4.2) node[midway, fill=white!0, rotate=-60] {$9.1\rm{~m}$};
    
    \draw[|<->|] (-24.22, -12.61) -- (-28.42, -4.53) node[midway, fill=white!5, rotate=30] {$9.1\rm{~m}$};
     
    \draw[ultra thick, -latex] (0+5, 0) -- (5,8);
    \draw[ultra thick] (4, 1) -- (6,1);
    \node[=none] at (5,10) {\Large North};

\end{tikzpicture}
    	}
    \caption{The red circle represents the commissioned dishes.}
    \end{subfigure}
     \caption[Currently commissioned dishes of TONE]{Arrangement of the commissioned dishes. Two dishes were damaged in a wind storm and one redundant dish which was outfitted with a slightly different analog chain was not chosen to be part of the final setup. The satellite imagery corresponds to the state of the array in the spring/summer of 2020.}
    \label{fig:dishes}
\end{figure}
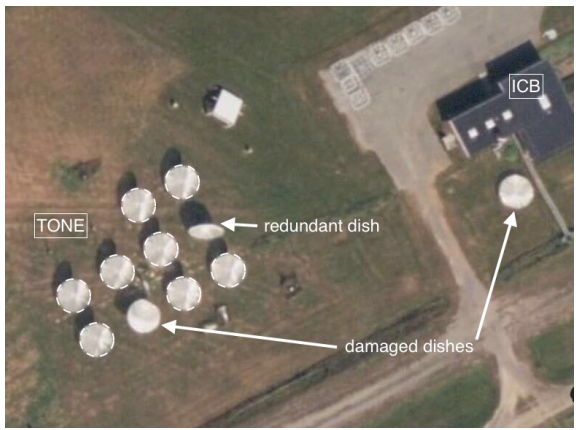
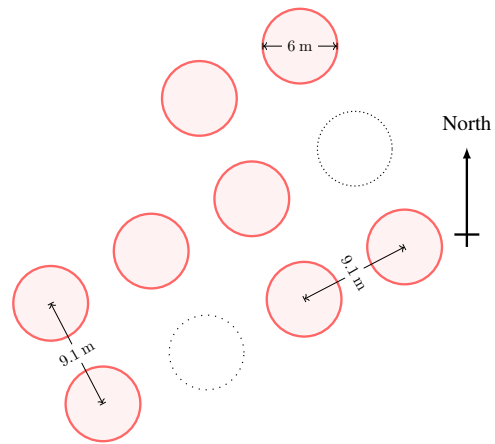

The array was designed to have $12\times$\SI{6}{\m} parabolic dishes arranged in a regular close-packed rectangular pattern. Each dish is an aluminum paraboloid reflector with an off-the-shelf steel frame which significantly reduces the cost of the array. The receiver is placed at the prime focus of each dish. We assembled 8 dishes in a rectangular grid, with the long axis oriented along a line \SI{60}{\degree} from North. A schematic drawing and the final layout of the dishes, as seen from satellite imagery, are shown in Figure~\ref{fig:dishes}. The dishes are pointed at a positive hour angle towards the declination of the Crab Nebula (hereafter, Taurus A) when it is at the meridian as viewed by CHIME. We use Taurus A as a primary calibrator to phase the antennas of TONE. The Crab pulsar's intermittent giant pulses are used as a calibrator over the long baseline for astrometric calibration.

\section{The Analog Chain}\label{sec:analog}

The analog chain (Figure~\ref{fig:analog}) consists of a dual-polarization cloverleaf-shaped dipole antenna with a full octave bandpass between \num{400} to \SI{800}{\mega\hertz} based on the same design as the CHIME antenna ~\cite{clover}. The active balun antenna has a low-noise amplifier (LNA) in the antenna stem. The amplified signal is passed via a coaxial cable to the radio frequency over fiber (RFoF) transmitter. The signal is sent via fiber in buried conduits to the interferometer control building into the digital backend where it is digitized.

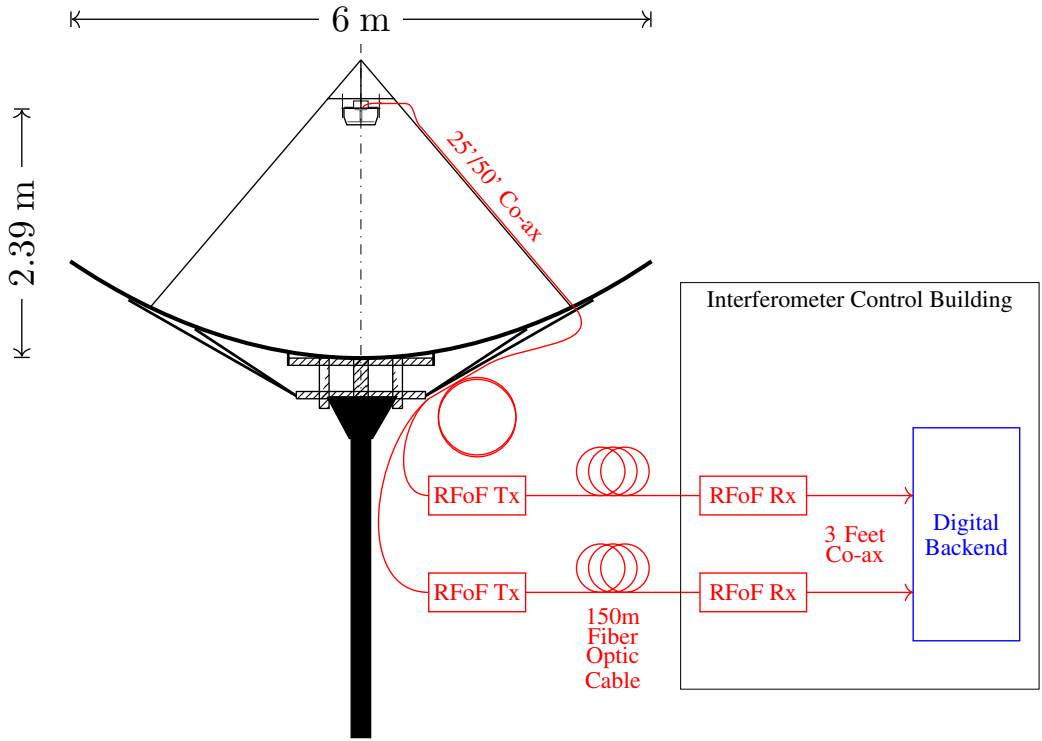
\begin{figure}[htp]
    \centering
    \resizebox{0.8\textwidth}{!}
    {    
    \begin{tikzpicture}[decoration={fiber, amplitude=5mm}]
\edef\feed{0.35}
\edef\D{6}
\edef\F{ \D * 0.43 }
\edef\baseh{0.42}
\coordinate (O) at (0, 0);

\draw[very thick] (O) parabola ++ (-\D / 2, 1);
\draw[very thick] (O) parabola ++ (\D / 2, 1);

\draw[fill=black] (-0.1,-0.5) -- ++ (.2,0) -- ++ (0, -\baseh - 3) -- ++ ( -.2,0) -| (-.1,-0.5);

\draw[fill=black] (-0.33,-0.4) -- ++ (.7,0) -- ++ (-.25, -\baseh - 0.01) -- ++ ( -.24,0) -- (-.35,-0.4);

\draw[thick] (-\D/8, -0.08) -- ++ (0, 0.15);
\draw[thick] (\D/8, -0.08) -- ++ (0, 0.15);           
\draw[pattern=north east lines] (-\D/8 , 0) -- (\D/8 , 0) -- ++ (0, -.075) -| (-\D/8, 0);
\draw[pattern=north east lines] (-\D/9 , -\baseh) -- (\D/9 , -\baseh) -- ++ (0, .075) -| (-\D/9, -\baseh);
\draw[pattern=north east lines] (-\D/14 , -\baseh-.1) -- (-\D/14 + 0.1 , -\baseh-.1) -- (-\D/14 + 0.1 , 0) (-\D/14, -\baseh-.1) -| (-\D/14, 0);
\draw[pattern=north east lines] (\D/14 , -\baseh-.1) -- (\D/14 - 0.1 , -\baseh-.1) -- (\D/14 - 0.1 , 0) (\D/14, -\baseh-.1) -| (\D/14, 0);
\draw[pattern=north east lines] (-0.075,0) -- ++ (.15,0) -- ++ (0, -\baseh - 0.1) -- ++ ( -.15,0) -| (-.075,0);

\draw[thick] (-\D/9 , -\baseh + 0.03) -- (-\D/3.5 , .3);
\draw[thick] (-\D/9 , -\baseh + 0.03) -- (-\D/2.5 , .6);

\draw[thick] (\D/9 , -\baseh + 0.03) -- (\D/3.5 , .3);
\draw[thick] (\D/9 , -\baseh + 0.03) -- (\D/2.5 , .6);

\draw[very thin] (-\feed/2, \F+0.011) -- ++  ( .1, 0 ) -- ++ (0, .065) -- ++ (.15,0) -- ++ (0, -0.065) --( \feed/2, \F+0.011);           
\draw[] (-\feed/2 , \F) -- ++ (0, -.09) -- ++  (+.03, -0.08) -- ++ (\feed-.06, 0) --  ++ (0.03,.09) -- (\feed/2, \F) -| (-\feed/2, \F);
\draw[thin] (-\feed/2-.025, \F) -- ( \feed/2+0.025, \F);           

\filldraw[gray, very thin] (-0.01, \F-0.013) -- ++ (.02,0) -- ++ (0, -0.125) -- ++ ( -.02,0) -| (-.01,\F);
\draw[thin, gray] (-.1385, \F-0.013-.125) -- ++ (0.277, 0);           
\draw[thin, gray] (-.065, \F-.013) -- ++ (0.13, 0); 

\draw[thin] (\D / 2.75, \F/5 ) -- (0, \F+.5);       
\draw[thin] (-\D / 2.75, \F/5 ) -- (0, \F+.5);       


\draw[thin] (0, \F) -- ++ (0, 0.5);           
\draw[very thin] (\feed/2 + .015, \F - 0.1) -- ++ (0, 0.25);
\draw[very thin] (-\feed/2 - .015, \F - 0.1) -- ++ (0, 0.25);           
\draw[thin] (-.35, \F+.1) -- ++ (.7, 0);           

\draw[dash dot, very thin] ($(O) + (0, -.75)$) node[below] {} -- ++ (0, \F + 1.5);

\draw[|<->|] ($(O) + (-\D / 2, 3.5)$) -- ($(O) + (\D / 2, 3.5)$) node[midway, fill=white] {$6\rm{~m}$};


 \draw[|<->|] ($(O) + (-.5 - \D / 2, 0)$) -- ++ (0, \F) node[midway, fill=white, rotate=90] {$2.39\rm{~m}$};



\node[draw=none, red, font=\fontsize{7}{0}\selectfont, text width=50,align=center, rotate = -49.5] at (1.4,\F-0.8) {25'/50' Co-ax};

\draw[thin, red](1.2,\F-3.2) circle (0.4);
\draw[thin, red](1.2,\F-3.18) circle (0.4);
\draw[thin, red] (0.02, \F) to[out=90,in=0] (0.2, \F+.05) -- (0.45, \F+.05)to[out=0,in=140](0.625, \F-0.2)--(2.24,\F-2.1) to[out=-60,in=30] (1.3,\F-2.65) -- (0.7,\F-3.) to[out=-140,in=-180] (0.7,\F-4.);
\draw[thin, red](0.7,\F-3.) to[out=-140,in=-180] (0.7,\F-5.);

\draw[thin,red] (0.7,\F-4.2) rectangle (1.7,\F-3.8) node[pos=.5, font=\fontsize{7}{0}\selectfont] {RFoF Tx};

\draw[thin,red] (0.7,\F-5.2) rectangle (1.7,\F-4.8) node[pos=.5, font=\fontsize{7}{0}\selectfont] {RFoF Tx};

\draw[thin, red, ->, decorate](1.7,\F-4) -- (3.5,\F-4);
\draw[thin, red, ->, decorate](1.7,\F-5) -- (3.5,\F-5);
\node[draw=none, red, font=\fontsize{7}{0}\selectfont,text width=30,align=center] at (2.6,\F-5.575) {150m Fiber Optic Cable };


\draw[thin,red] (3.5,\F-3.8) rectangle (4.6,\F-4.2) node[pos=.5, font=\fontsize{7}{0}\selectfont] {RFoF Rx};

\draw[thin,red] (3.5,\F-4.8) rectangle (4.6,\F-5.2) node[pos=.5, font=\fontsize{7}{0}\selectfont] {RFoF Rx};

\draw[thin, red, ->](4.6,\F-4) -- (5.7,\F-4);
\draw[thin, red, ->](4.6,\F-5) -- (5.7,\F-5);
\node[draw=none, red, font=\fontsize{7}{0}\selectfont, text width=20,align=center] at (5.1,\F-4.5) {3 Feet Co-ax};

\draw[thin,blue] (6.8,\F-3.3) rectangle (5.7,\F-5.5) node[pos=.5, font=\fontsize{7}{0}\selectfont, text width=35,align=center] {Digital Backend};

\draw[very thin] (7,\F-1.8) rectangle (3.3,\F-6);
\node[draw=none, font=\fontsize{7}{0}\selectfont] at (5.15,\F-2) {Interferometer Control Building};

\end{tikzpicture}
    }

    \caption[The analog chain of a single dish of TONE.]{The analog chain of a single dish of TONE. The signal collected by the active feeds from each polarization is sent over 
    \num{25} or \num{50} feet of coaxial cables (not all cables are the same length for all the dishes -- some are \num{25} feet and some \num{50} feet). The signal from the coaxial cables is sent over \SI{150}{\meter} of fiber optic cable to an electromagnetic-compliant rack inside the interferometer control building using a custom RFoF system. The receiver converts the light back to an electrical signal that is then digitized by the ICE boards.}
    \label{fig:analog}
\end{figure}

\subsection{Cloverleaf Antenna \& Low Noise Amplifier}

We use a dual-polarization cloverleaf feed (Figure~\ref{fig:clover}) based on the design that was developed for CHIME~\cite{clover} and which was selected for HIRAX~\cite{hirax2, hiraxantenna}. It is an active feed that consists of a balun that uses an Avago MGA-16116 dual amplifier, and the difference between the outputs is amplified using a Mini-Circuits PSA4-5043+ amplifier. Each feed is mounted inside a cylindrical can, which circularizes the beam and helps reduce crosstalk between dishes. Each antenna-can assembly is mounted at the focus on struts extending from the dish and a custom pyramid structure. The active antennas are powered by two central power supply units supplying \SI{\sim7}{\volt} directly over copper electrical cables. The feeds are protected from the elements by custom vacuum-formed plastic covers and neoprene seals. 

\begin{figure}[htp]
 \centering
\includegraphics[width=0.5\textwidth,angle=-90,keepaspectratio]{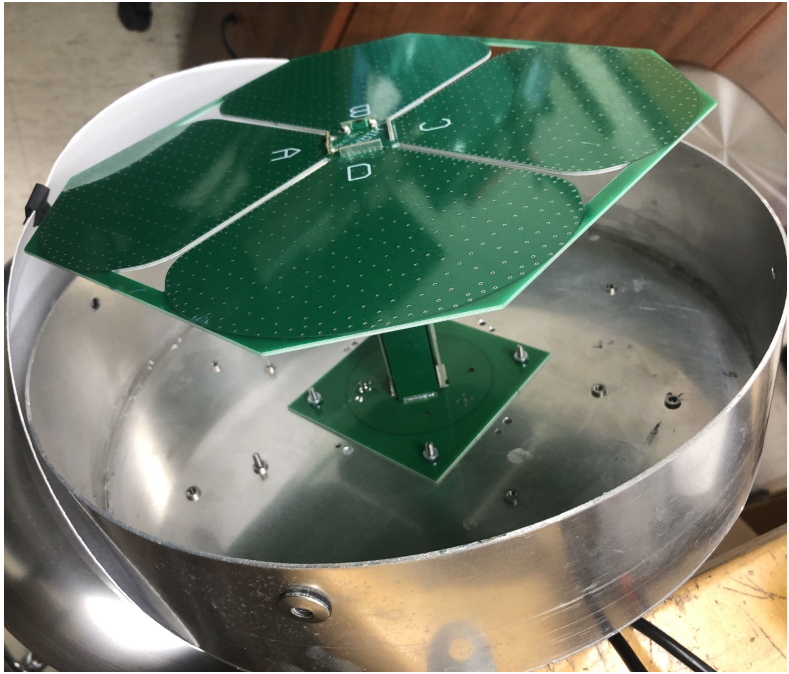}
  \caption[The TONE Feed]{The feed with the cloverleaf antenna. The polarizations are aligned with the printed letters on the feed i.e., rotated \SI{45}{\degree} from the slots. The LNA is in the stem.}
  \label{fig:clover}
\end{figure}

\subsection{Radio Frequency over Fiber (RFoF) system}

\begin{figure}[htp]
  \centering
  \includegraphics[width=0.5\textwidth,angle=-90,keepaspectratio]{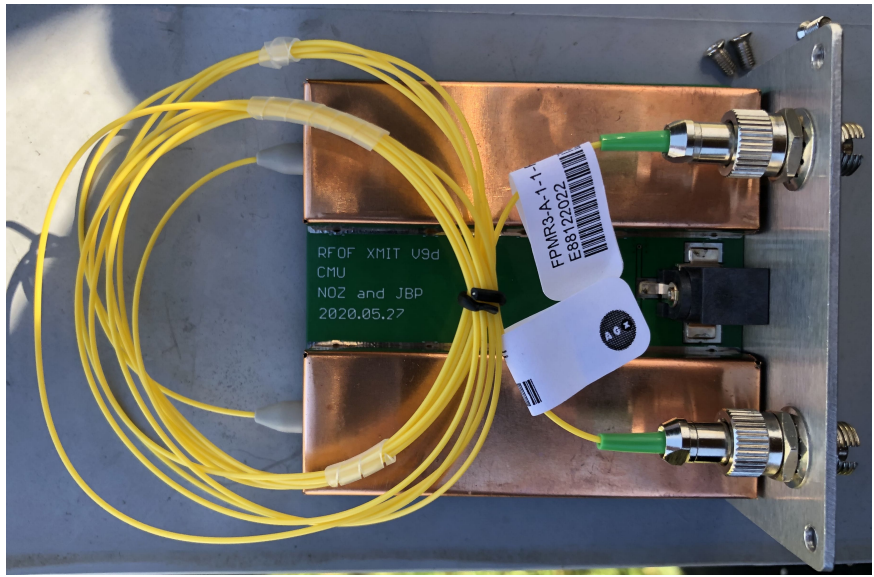}
  \caption[The RFoF system]{The RFoF transmitter, as seen inside its electromagnetically compliant enclosure. }
  \label{fig:rfof}
\end{figure}

Signals received by each polarization of the feed are sent over coaxial cables to a junction box at the pole below every dish. This is fed into an RFoF transmitter module (Figure~\ref{fig:rfof}). It is then band-limited to \SIrange{400}{800}{\mega\hertz} and passed through an amplification stage before being intensity-modulated on an optical carrier. \SI{150}{\meter} of optical fibers carry these signals to the digital backend server rack, where RFoF receivers convert the signals back into electrical voltages. The RF signals are amplified and filtered again before being passed to the ICE boards (see \S\ref{subsec:ice_board}). The transmitter contains a laser diode (AGX Technologies, FPMR3 series) that is intensity-modulated by the incoming RF signal, and the receiver contains a photodetector (AGX Technologies, PPDD series) that converts the transmitted optical signal into radio frequency voltages. The RFoF transmitter and receiver design are based on technology that was developed for CHIME~\cite{rfof}.

\section{Digital system}\label{sec:digital}

Much like CHIME~\cite{chimeoverview, chimesystem}, the TONE backend consists of an `F' engine and a high-bandwidth data recorder node. In the ``F'' engine, called the ICE Boards (see \S\ref{subsec:ice_board}), a total of \num{16} wideband analog signals from the analog chain in the \SIrange{400}{800}{\mega\hertz} frequency band are alias-sampled at \num{800} Msps into the second Nyquist zone by the analog to digital converters (ADCs) on the mezzanines of the ICE boards.

\begin{figure}[htp]
    \centering
    \includegraphics[]{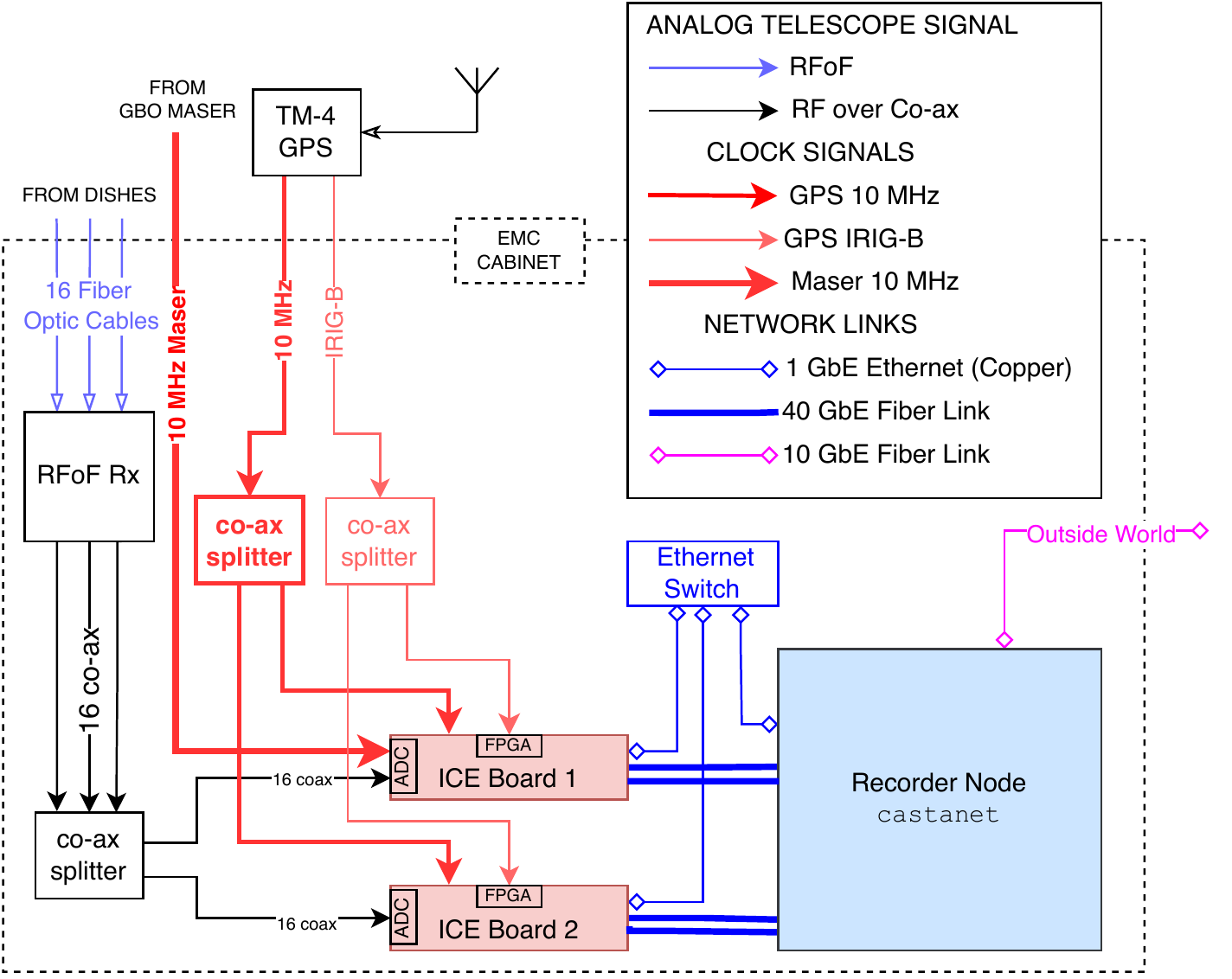}
    \caption[Digital backend Diagram]{The flow diagram shows the entire system as present in the electromagnetically compliant cabinet. The two redundant ICE boards can be used interchangeably. ICE Board 1 is used as an F engine for CHIME-TONE VLBI, digitizing the sky inputs and maser input, and forwarding all of the channelized voltages to the recorder node over a high-bandwidth network interface. ICE Board 2 is configured as a full FX correlator for operating TONE independently of CHIME, computing and sending full visibilities to the recorder node via the switch.}
    \label{fig:tonedigitalchain}
\end{figure}

\subsection{ICE Boards}\label{subsec:ice_board}

The F-engine for TONE consists of two ICE boards~\cite{ice}. Each board is a custom FPGA motherboard that makes use of a Xilinx Kintex-7 FPGA and ARM-based co-processor. Each board processes \num{16} digitized inputs from \num{8} dual-polarization feeds. Since TONE consists of $\leq 8$ antennas operating in tandem, each board is fed an identical copy of the analog signals from the RFoF receiver that have been split with simple resistive coaxial splitters. The two boards both perform channelization of the data but serve different purposes downstream. ICE Board 1 is used for VLBI with CHIME while ICE Board 2 is used to operate TONE as an independent instrument. Both ICE boards are controlled by a custom Python software, \texttt{pychfpga}\footnote{\url{https://bitbucket.org/winterlandcosmology/pychfpga/}}, which is used to communicate with and program the FPGA and the ARM co-processor. The ICE boards are both synchronized using a \SI{10}{\mega\hertz} signal. Absolute time is provided to each board in the IRIG--B format \footnote{\url{https://www.wsmr.army.mil/RCCsite/Documents/200-16_IRIG_Serial_Time_Code_Formats/200-16_IRIG_Serial_Time_Code_Formats.pdf}} from a TM--4 GPS receiver \footnote{A modified TM--4D receiver with the time distribution system disconnected and removed~\cite{tm-4d} \url{https://www.spectruminstruments.net/products/tm4d/tm4d.html}}. 

The ICE boards receive and digitize the 16 incoming signals from the RFoF system at a rate of \num{800}~Msps with a dynamic range of 8 bits over the full \SI{400}{\mega\hertz} of bandwidth. Each board, when in use, saves a $\SI{2.56}{\micro\second}$-snapshot of this ``pre-channelization'' voltage data from each input every second. These snapshots are sent over the gigabit network and written to disk; we refer to this data stream as raw ADC data henceforth. 

The data are then channelized by a polyphase filter bank and an FFT-based pipeline, producing complex voltages in 1024 frequency channels (each \SI{390.625}{\kilo\hertz} wide) with \SI{2.56}{\micro\s} time resolution. We refer to these channelized voltages as our ``baseband'' data. After the channelization, each FPGA performs a real-time transpose, or ``corner-turn,'' in so-called \texttt{shuffle16} mode. This arranges the 16 outgoing data streams such that each network interface of the recorder node receives data from each of the 16 inputs over a subset of the total bandwidth. The network interface consists of two \num{40} GbE QSFP+ fiber links over $8\times10$ gigabit lanes sending \num{128} frequency channels in each lane. 

ICE Board 1 channelizes data and passes them along to the baseband buffer and recorder for VLBI. It also receives a \SI{10}{\mega\hertz} sine wave distributed from the Microsemi MHM 2010 Active Hydrogen Maser at Green Bank. The \SI{10}{\mega\hertz} sine wave is fed to one of the inputs of ICE Board 1 to measure and correct for jitter in the TM--4 clock signal. ICE Board 2 receives no maser signal and performs a full correlation and integration. This is particularly useful for observing transits of bright sources which are used as calibration sources for TONE. In this so-called ``correlator mode'', all visibilities are sent over a gigabit link to the recorder node~\cite{ice2}.

\subsection{High-bandwidth VLBI Recorder}
In CHIME, the channelized, corner-turned baseband data is passed along via corner-turn to an X-engine which computes visibilities in real-time. The TONE system design was strictly constrained by the science case of saving baseband data to cross-correlate for VLBI localization. As seen in Figure~\ref{fig:tonedigitalchain} the data from each ICE board is sent via \num{2x40} GbE links to \texttt{castanet}, our high bandwidth recorder node. In \texttt{castanet}, baseband data are continuously written in parallel to a bank of high-density memory modules, which host a \SI{\sim40}{\s} long ring buffer. Upon a ``trigger,'' a \SI{500}{\milli\second} segment of the memory buffer, sliced over time and frequency to follow the dispersive sweep, is saved to disk. The slicing and readout algorithms are implemented in \texttt{kotekan} \footnote{\url{https://github.com/kotekan/kotekan}}\cite{kotekan}: a high-performance real-time data processing software framework designed for modern radio telescopes. The hardware components of the high-bandwidth VLBI recorder are summarized in Table \ref{tab:specs}.

\begin{table}[htb]
\begin{tabularx}{\textwidth}{p{0.175\linewidth} | p{0.35\linewidth}|p{0.35\linewidth}}
\hline
\textbf{Parts} & \textbf{Part Number}                     & \textbf{Specifications}                                           \\ \hline
Motherboard    & TYAN Tempest EX S7100-EX                 & $4 \times$PCIeX16, $3\times$PCIeX8, 2 sockets                        \\
CPU            & $1\times$Xeon Silver 4116 & 12 cores (hyperthreaded) $\times$2.1 GHz           \\
Memory         & $4\times$HYNIX HMAA8GR7A2R4N-VN                 & 128 GB                                                            \\
Network        & $2\times$Silicom PE 31640G2QI71/QX4             & $2\times4\times$10 GbE\\ \hline
\end{tabularx}
\caption[TONE backend system]{\label{tab:specs} Components of the High-bandwidth VLBI Recorder. The recorder node was able to accommodate 190 gigabits/s of bandwidth corresponding to the full bandwidth of all the inputs from both ICE boards, but the final configuration of TONE did not use this full bandwidth.}
\end{table}

\begin{figure}[htp]
\centering
  \includegraphics[width = 0.5\textwidth]{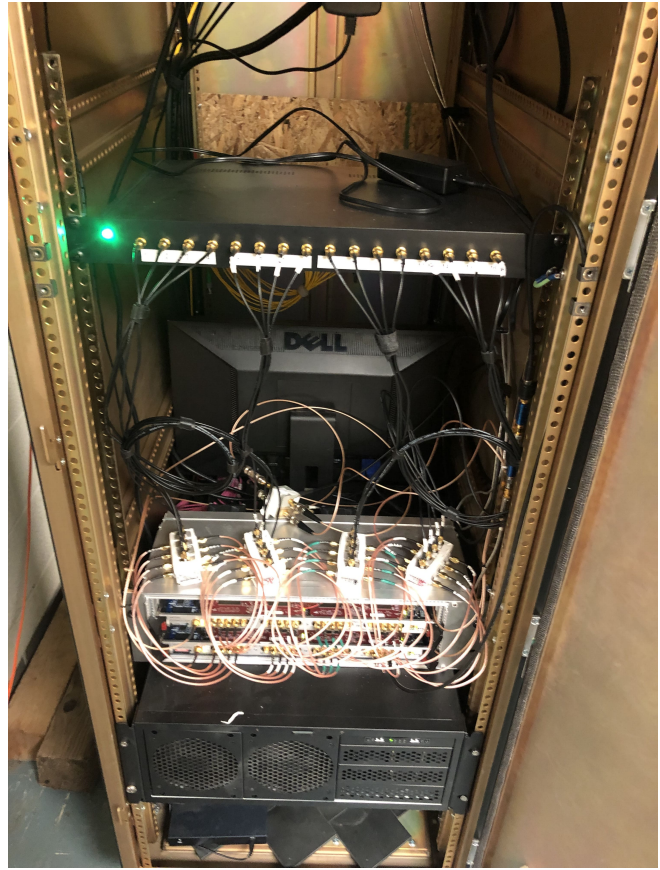}
  \caption[Digital back-end of TONE]{A picture of the digital backend inside the electromagnetically-compliant rack. \textit{From the top:} The RFoF receiver, the ICE boards, and the Recorder node. The base of the rack holds power supplies for the ICE boards and a gigabit switch for the internal network.}
\end{figure}

\subsection{VLBI Triggering System}

To facilitate concurrent observations, an asynchronous server called \texttt{frb-outrigger} was developed. A reverse ssh tunnel is set up to the CHIME network.  A ``heartbeat'' signal from \texttt{frb-outrigger} is sent to CHIME/FRB's backend informing that the outrigger is online as well as the relevant CHIME/FRB sky beams to monitor for FRB detection. On detection of a signal by the CHIME/FRB real-time pipeline, a trigger is sent over the reverse ssh tunnel. The script extracts the time of arrival of the pulse, the DM, and the DM uncertainty. This information is sent to the \texttt{kotekan} baseband endpoint which saves a  \SI{500}{\milli\second} slice of baseband data centered around the frequency dispersed pulse to disk. Additionally, the trigger script gathers information from the incoming trigger and the path to the raw ADC data corresponding to the time of the trigger. The latency of the system from data readout to disk against the trigger time is an average of \SI{\sim14}{\second}. Triggers are also sent for Crab pulsar giant pulses \cite{2015MNRAS.446..857L} detected by CHIME/FRB in the TONE field-of-view. We record Crab pulsar giant pulse data triggered with a detection S/N greater than \num{40} (near CHIME's zenith) with a duty cycle of \SI{1}{\percent}. This prevents overwhelming the baseband readout system with thousands of Crab pulsar giant pulsars and results in a Crab GP data acquisition rate of about once per day.

\begin{figure}[htp]
    \centering
    \includegraphics[width=.75\textwidth, keepaspectratio]{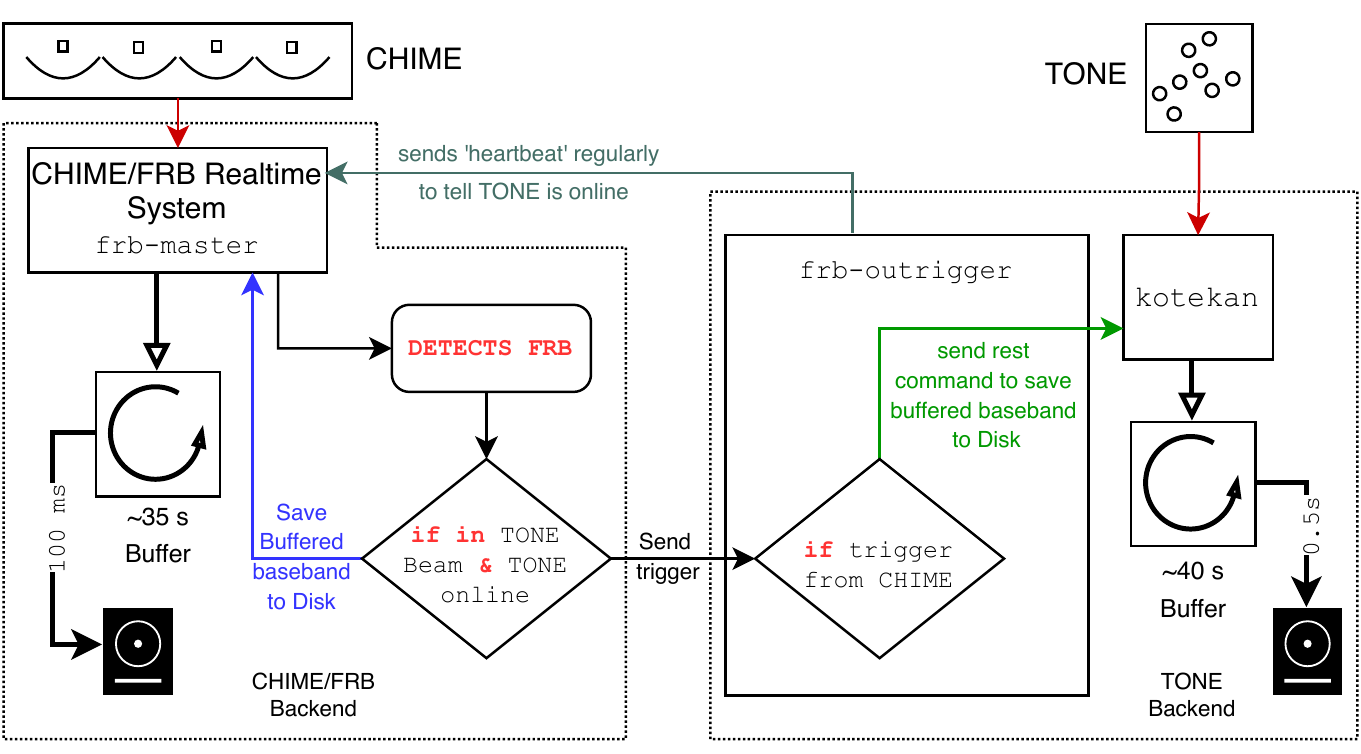}
    \caption[Triggering Infrastructure]{A diagram describing the triggering infrastructure.}
    \label{fig:tonetrigger}
\end{figure}

\section{Performance of the Telescope}\label{sec:performance}

\subsection{Analog Chain}
\begin{figure}[htp]
\centering
\includegraphics[width=0.5\textwidth]{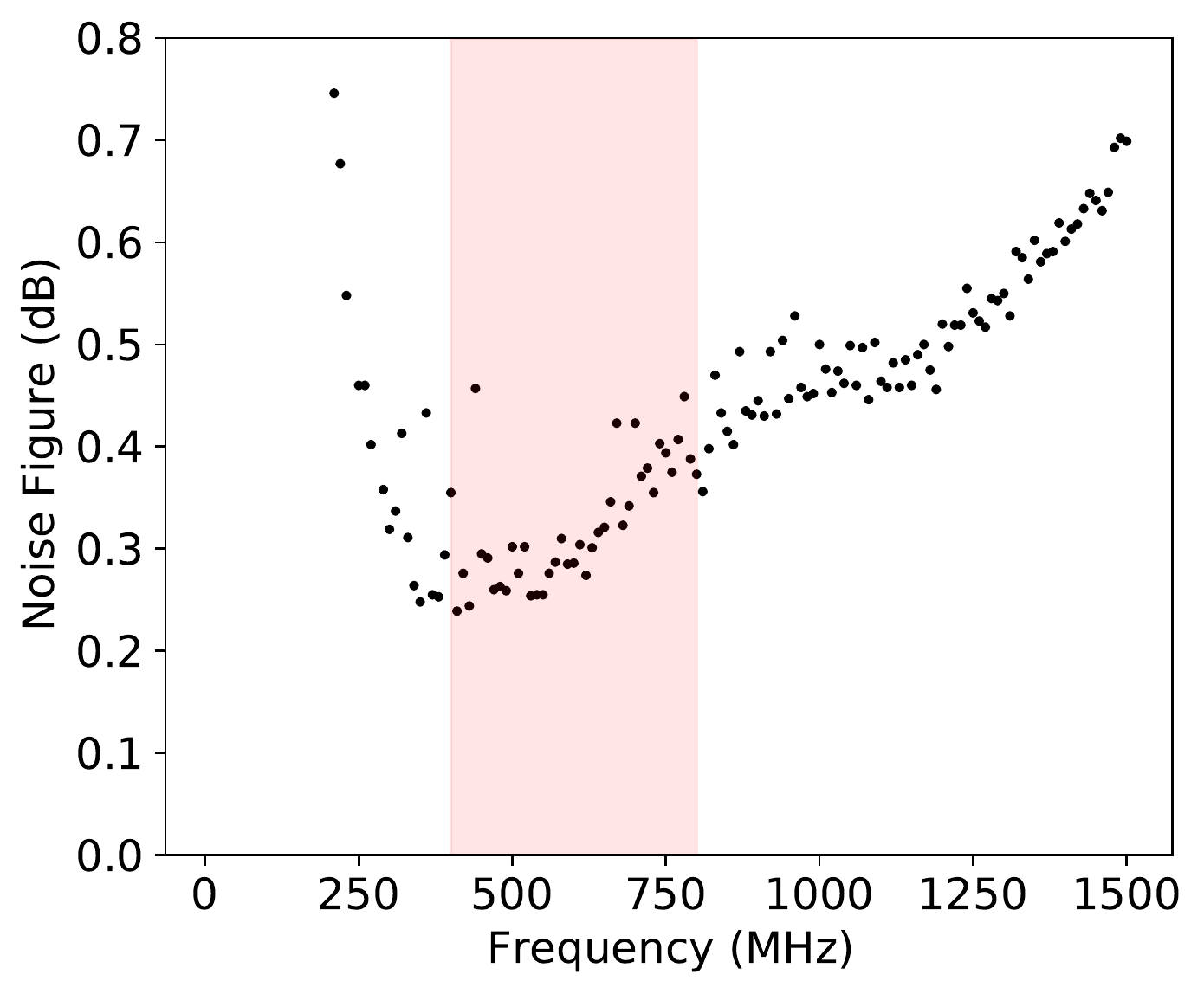}
\caption[TONE LNA characteristics: Noise Figure]{Noise Figure of the LNA. The red region shows the frequency of operation of TONE.}
\label{fig:nflna}
\end{figure}

\begin{figure}[htp]
\centering
\includegraphics[width=0.4\textwidth, angle=-90]{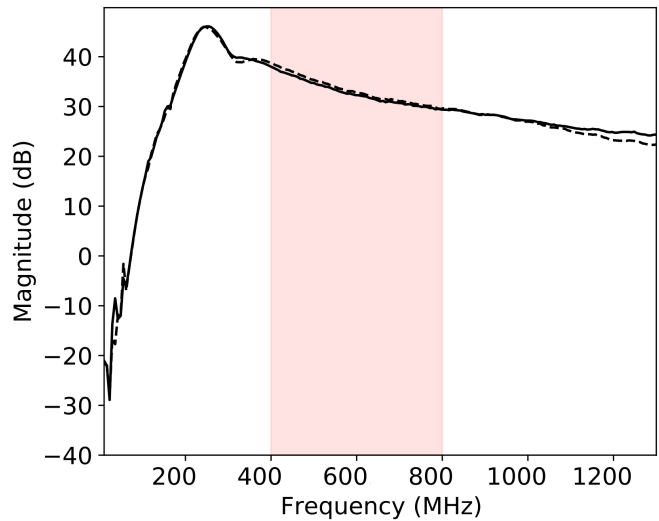}
\caption[TONE LNA characteristics: LNA gains]{The gain of the LNA. The solid and dashed lines show the gains of the two channels of the LNA.}
\label{fig:s21lna}
\end{figure}

\begin{figure}[htp]
  \centering
  \includegraphics[width=0.5\textwidth]{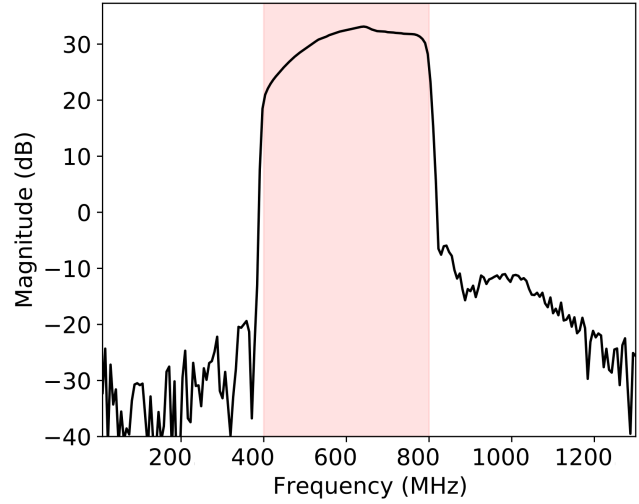}
  \caption[RFoF system gain]{ The gain in dB of the RFoF system. The red region shows the frequency of operation of TONE.}
  \label{fig:s21rfof}
\end{figure}

\begin{figure}[htp]
  \centering
  \includegraphics[width=0.5\textwidth]{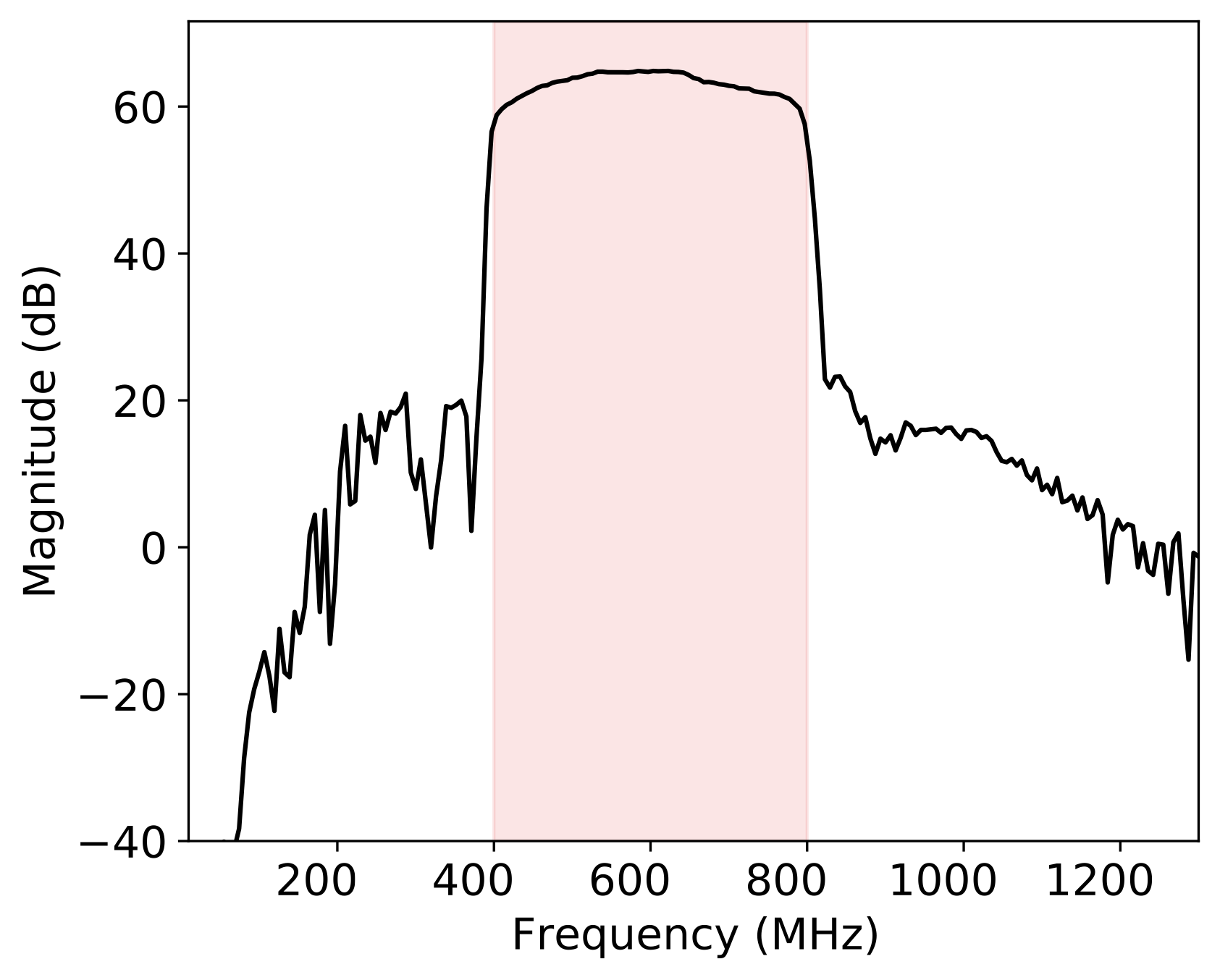}
  \caption[Analog chain gain]{The gain in dB of the entire analog chain from the LNA input to the RFoF receiver output. The red region shows the frequency of operation of TONE.}
  \label{fig:s21all}
\end{figure}

The characteristics of the analog chain were measured in the lab. We measured the gain and noise figures of the low-noise amplifier (LNA). The noise figure is the figure of merit of the degradation of the signal-to-noise ratio (lower is better); both the noise figure and gain are shown in Figure~\ref{fig:nflna} and Figure~\ref{fig:s21lna} respectively. The RFoF chain, the transmitter, and the receiver gain are shown in Figure~\ref{fig:s21rfof}. The gain (or $S_{21}$ parameter) of the entire analog chain from the LNA to the RFoF chain is shown in Figure~\ref{fig:s21all}.

\begin{figure}[htp]
  \includegraphics[width=\linewidth]{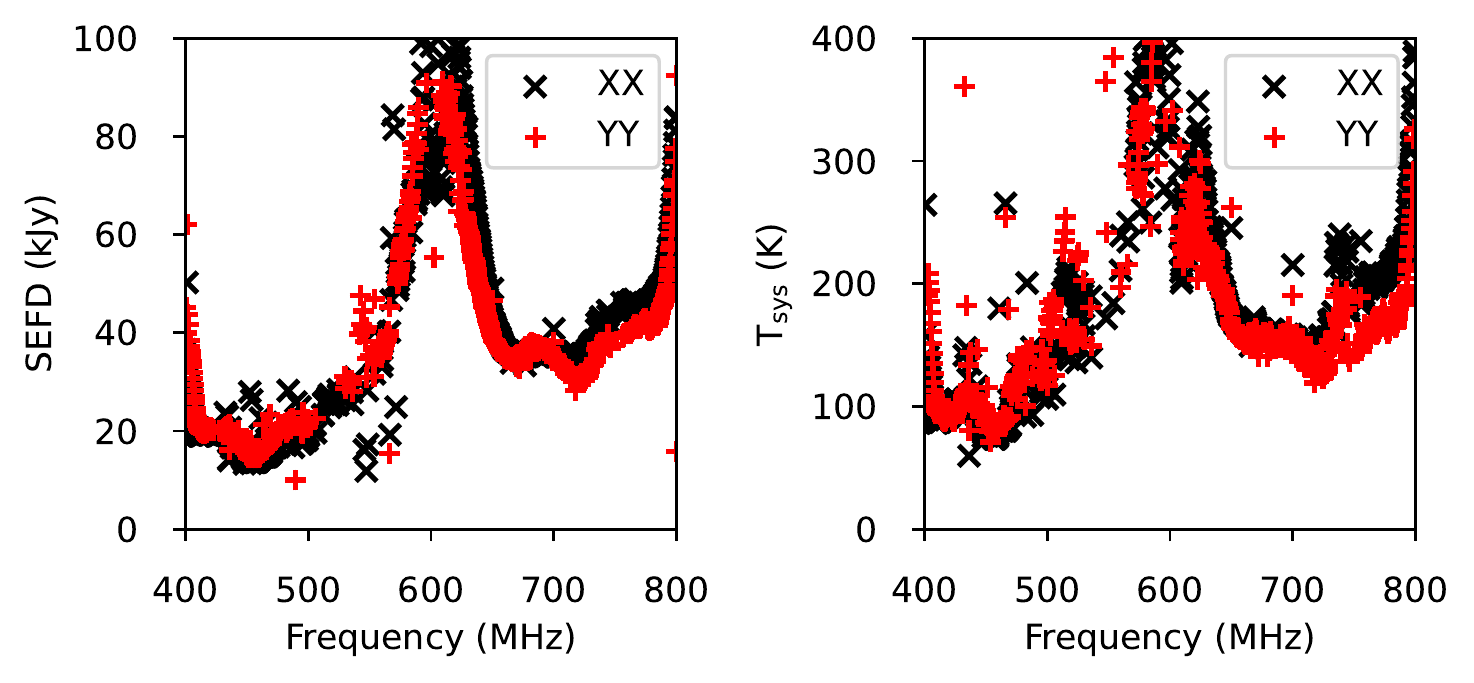}
  \label{fig:sefd}
\caption[System equivalent flux density (SEFD) and System Temperature]{\textit{Left:} System equivalent flux density (SEFD) of a single dish. \textit{Right:} System temperature of a single dish. The hump of higher SEFD and system temperature between $\sim$\SIrange{550}{650}{\mega\hertz} is partially caused by the impedance mismatch and the signal multipath effect of reflections between the feed ground plane and dish. }
\label{fig:test}
\end{figure}

\begin{figure}[htp]
\centering
  \centering
  \includegraphics[width=\linewidth]{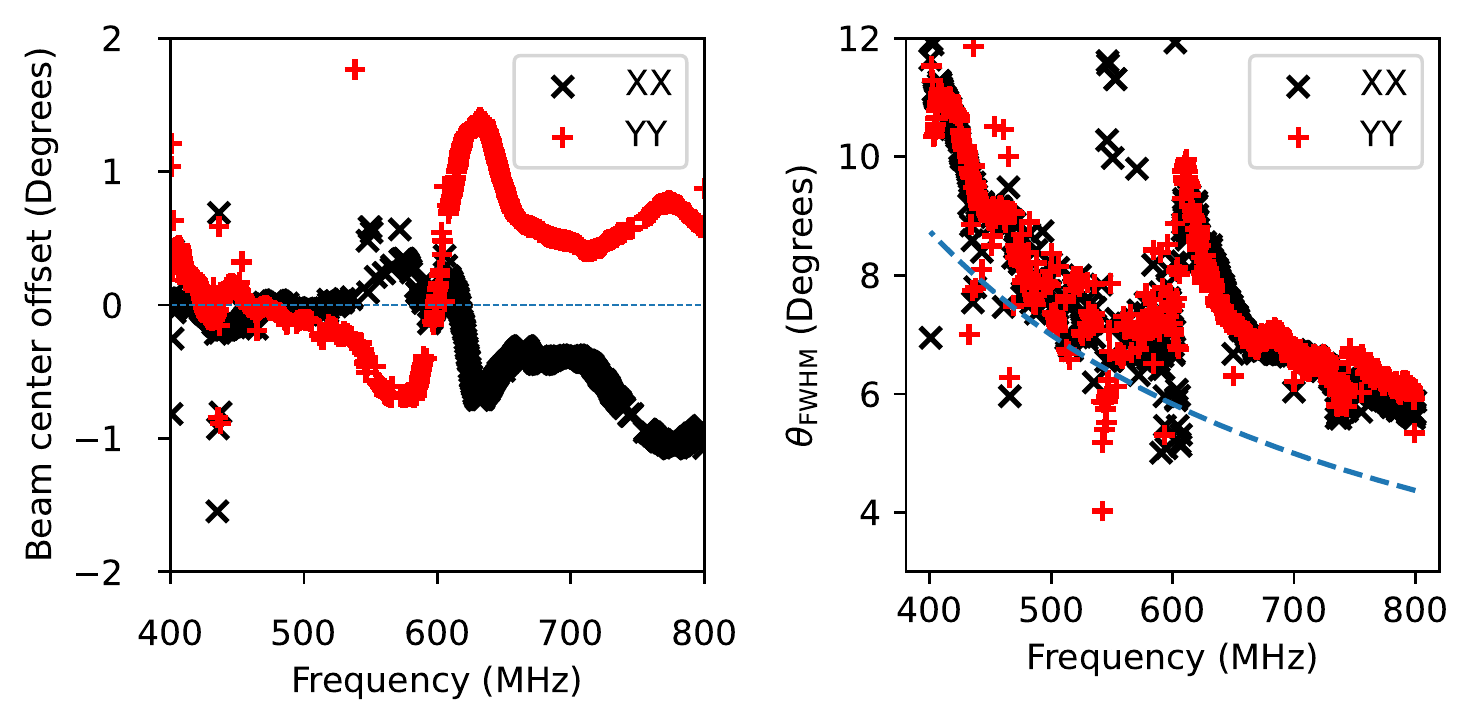}
\caption[TONE beam properties]{\textit{Left:} The directional offset of the beam bore-sight from the expected pointing. While there is a small, frequency-dependent systematic offset between the two polarization, it is much smaller than the beam width. \textit{Right:} the full-width half max (FWHM) of the primary beam of a single dish. The blue dashed line corresponds to the expected FWHM of a parabolic dish. The hump of higher FWHM between $\sim$\SIrange{550}{650}{\mega\hertz} is partially caused by the impedance mismatch and the signal multipath effect of reflections between the feed ground plane and dish.}
\label{fig:beam}
\end{figure}

We assess the performance of the system using on-sky data. We characterize many aspects of the TONE analog chain by acquiring visibility data of the Taurus A transit in the correlator mode from the ICE board and analyze the cross-correlations. We assume that the dishes and receivers are roughly similar and the beams have circular Gaussian symmetry. The characteristics of each dish can be measured in the visibilities between two dishes if we assume the dishes are similar. We use the visibility from the same polarization of two feeds that are far away from each other (separated by at least one dish between them) to prevent the effects of cross-talk between the feeds and reflections. At each transit, at each frequency, the cross-correlation between the signals in the visibility data is fit to a Gaussian profile (with free parameters of offset, amplitude, mean, and width). The amplitude is then calibrated to the flux from Taurus A~\cite{flux}. The calibration factor is then scaled to the noise, i.e. the off-source sky, in the autocorrelation visibilities to compute the system equivalent flux density (SEFD) is shown in Figure~\ref{fig:test}. The width of the Gaussian is used to compute the full-width half maximum ($\fwhm$) at each frequency shown in Figure~\ref{fig:beam}. The solid angle of a Gaussian beam, $\Omega \approx 1.113 \fwhm^2$ is then used to compute the system temperature at each frequency $\Tsys = ({10^{-26} \lambda^2})/({2 \kb \Omega}) \times \mathrm{SEFD}$ [Jy] shown in Figure~\ref{fig:test}. The mean of the fitted Gaussian corresponds to the time of the transit through the boresight of the beam. We can thus estimate the offset from the correct pointing at each frequency, as seen in Figure~\ref{fig:beam}. The pointing of the dishes' structure is within a degree of expected and is limited by the accuracy of compasses used to manually point the dishes and feeds. These values of the system temperature, beam offsets, FWHMs, and SEFDs were computed for several pairs of all active dishes in the data. They were found to be similar for every dish. This measurement of the system temperature is conservative since the data are affected by the Galaxy, spillover, the ground, and pointing offset.

\subsection{Timing}
\begin{figure}[htp]
 \centering \includegraphics[width=\linewidth]{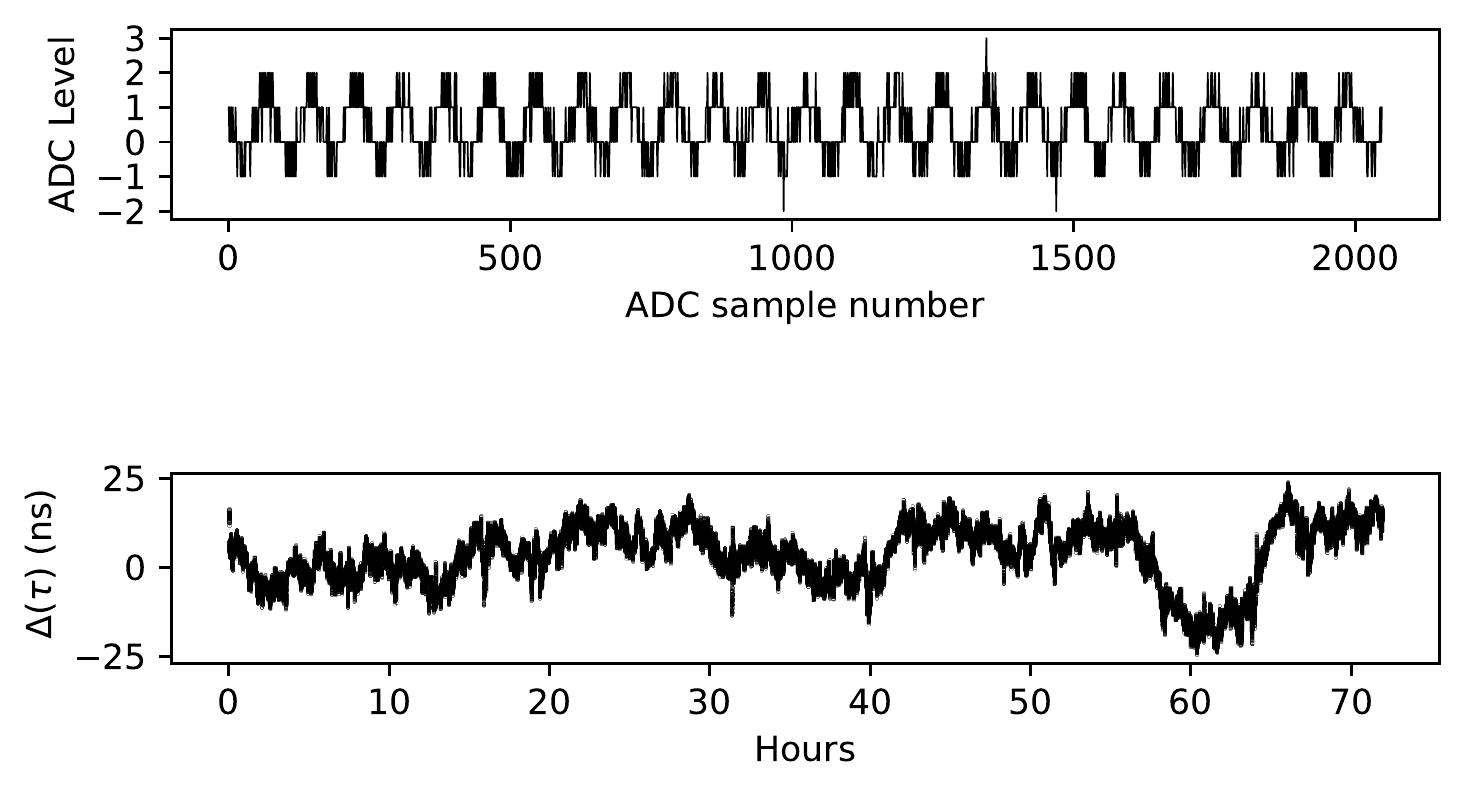}
  \caption[The GPS clock jitter]{\textit{Top:} A single frame raw ADC data of the quantized \SI{10}{\mega\hertz} maser signal as recorded by the ICE board ADC sampled at \SI{800}{\MSPS}. \textit{Bottom:} The deviations (jitter) of the GPS \SI{10}{\mega\hertz} clock as measured against the maser \SI{10}{\mega\hertz} clock.}
  \label{fig:gpsvmaser}
\hfill
\end{figure}

\begin{figure}[htp]
  \centering
\includegraphics[width=\linewidth]{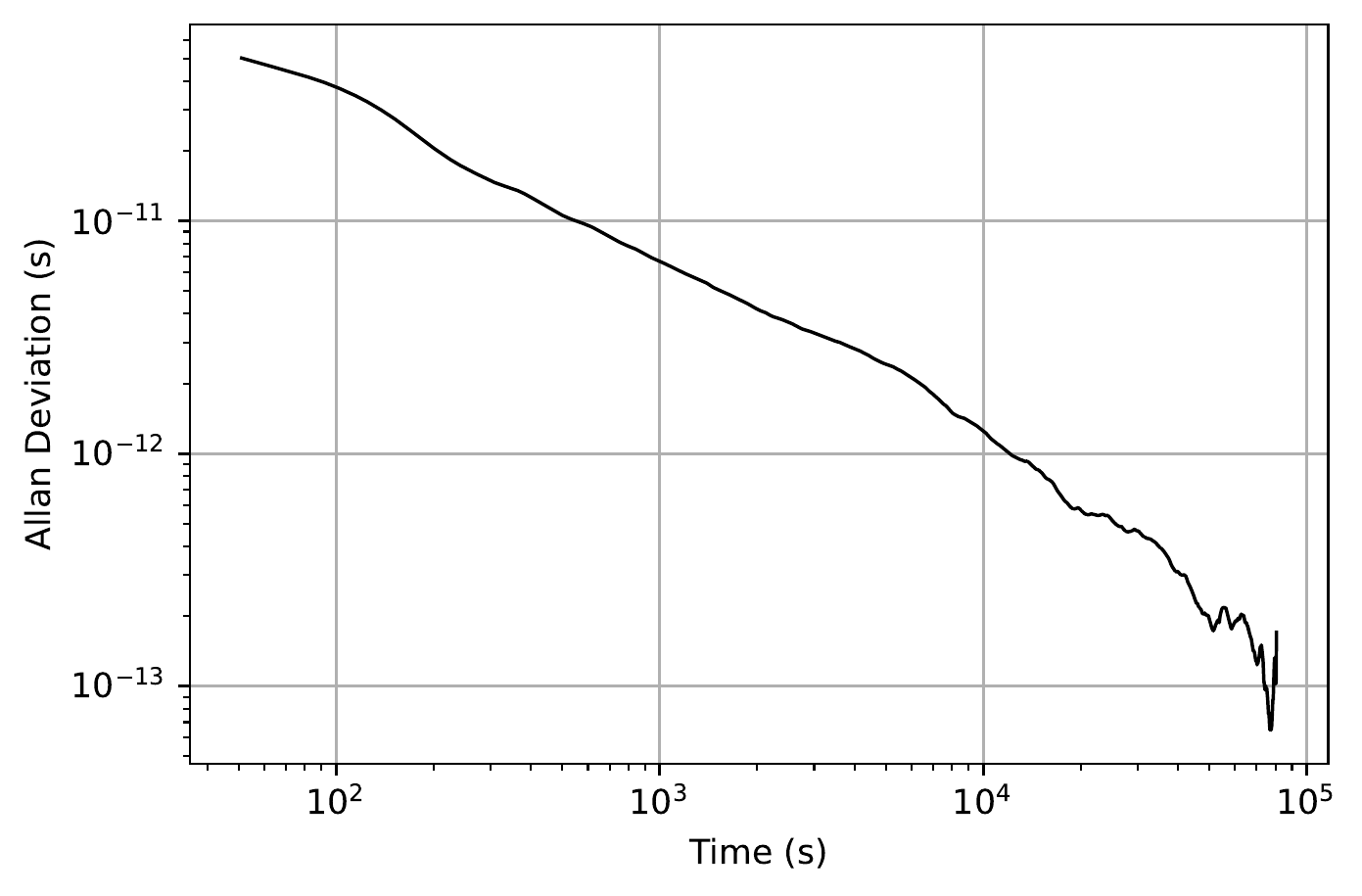}
  \caption[Allan Deviation of the GPS clock]{The measured overlapping Allan deviation of the TM--4 GPS unit. On the timescales of $\Delta t \sim \SI{e3}{\second}$, the clock stability corresponds to a root mean square (RMS) timing error of \SI{\sim6}{\nano\second} under the assumption of Gaussian timing fluctuations, and is consistent with Figure 1 of~\citet{2022AJ....163...48M}.}
  \label{fig:adev}
\end{figure}

Our science goals require us to maintain a highly-stable digital sampling cadence and a precise sense of absolute time~\cite{timingoutriggers} over timescales of months. This requires active stabilization of the GPS \SI{10}{\mega\hertz} master clock, which in turn introduces jitter on short timescales as seen in Figure~\ref{fig:gpsvmaser}. Information about this jitter can be reconstructed on one-second timescales using the raw ADC snapshots sent by ICE Board 1 since one of the ADC inputs is fed a copy of the GBO maser signal. The Fourier transform of the time stream of the maser input is taken out of raw ADC data; the phase of the alias-sampled \SI{10}{\mega\hertz} channel encodes the frequency offset between the maser and the GPS clock. The overlapping Allan deviation~\cite{allan1966statistics} of the GPS clock measured against the reference maser is measured and shown in Figure~\ref{fig:adev}. This Allan deviation can be attributed to the GPS clock under the assumption that the reference signal as distributed to TONE from the active hydrogen maser does not change significantly over timescales of several hours. 

\subsection{Array Calibration}\label{sec:cal}
As a check of the stability of the full signal chain, we measure the phases and amplitudes of each antenna input referenced to a fixed input. These so-called complex gains include static effects like differing cable lengths and phase delays from analog electronics, as well as time-dependent factors like temperature-induced fluctuations in the transmission cables.

Algorithms and infrastructure to calculate these complex gains, called the ``$N^2$-gain calibration'', have already been developed for CHIME~\cite{2014SPIE.9145E..22B,2015arXiv150306189R,2021ApJ...910..147M}. We generalize several software frameworks to use the same basic $N^2$-gain calibration algorithms for TONE data. From a sequence of $\SI{500}{\milli\second}$ baseband data collected when Taurus A is in the TONE primary beam, we calculate $N^2$ visibility matrices in each frequency channel for each of the two groups of orthogonal polarisations. This gives a total of 1024 matrices of dimension $8 \times 8$ for each of the two polarization groups. We clean each matrix by zeroing its diagonal matrix and deleting rows/columns corresponding to antennas whose performance is anomalous. We fringestop the cleaned visibility matrix to correct for Taurus A being off-zenith and perform an eigendecomposition of the fringestopped matrix. The components of the dominant eigenvector are the complex antenna gains for that frequency channel/polarization group.

\begin{figure}[htp]
  \centering
  \includegraphics[width=\linewidth]{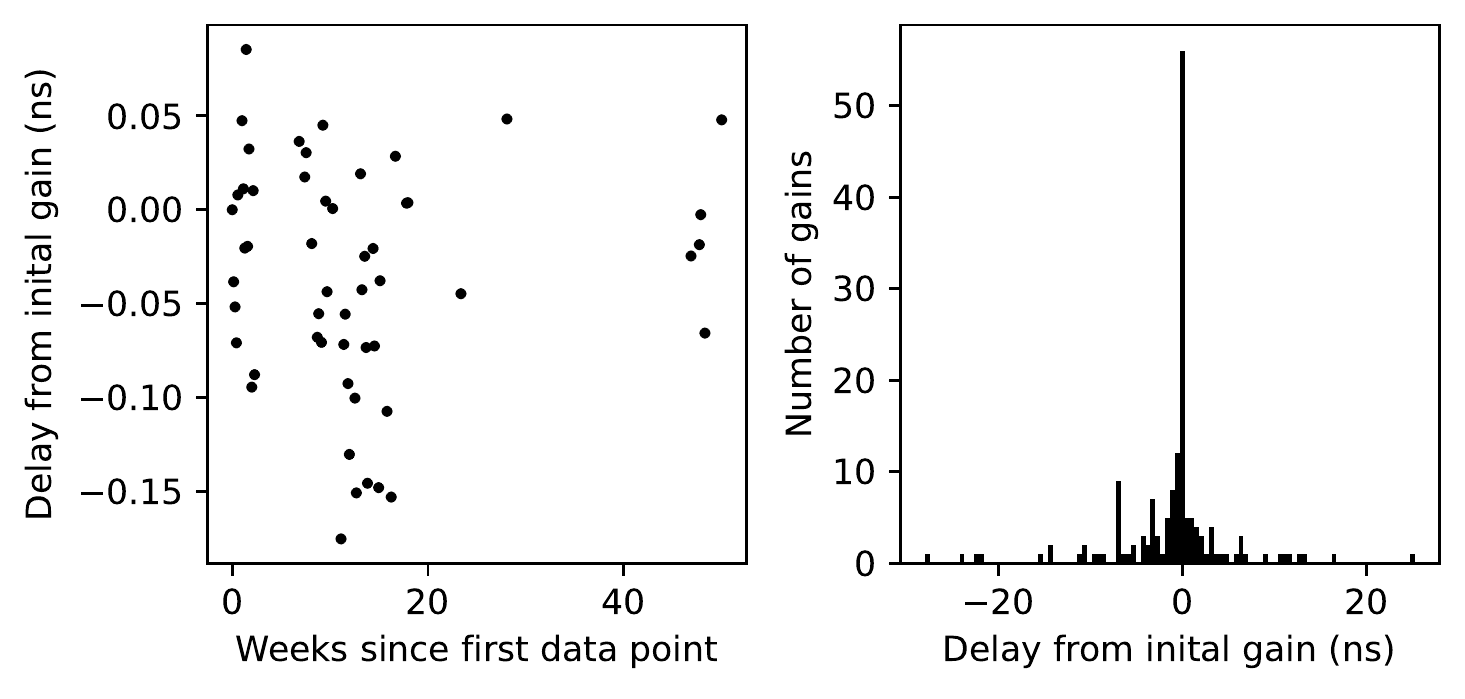}
\caption[Instrumental delays]{The characteristic instrumental delay of a typical analog signal from one polarization of one antenna. \textit{Left}: delays relative to the first day after removing outliers corresponding to bad gains solutions due to RFI, bad weather, and the sun in the beam. \textit{Right}: We plot the full histogram of instrumental delays; bad gains solutions correspond to histogram entries at large delay values.}
\label{fig:cabledelays}
\end{figure}

These complex gains essentially represent the phasing that needs to be applied to data from the dishes to form a beam directly at the zenith i.e. compensating for the instrumental delays for each dish. Thus, we unwrapped the phase of the complex gains as a function of frequency by fitting for the delays $\tau$ in $\phi = 2 \pi \nu_i \tau$, where $\nu_i$ for $i = 0,1,\ldots,1023$ refer to the central frequency of each channel. Computing the gains for all the baseband data recorded for Taurus A, we get the delays shown in Figure~\ref{fig:cabledelays} referenced to the delay from the first day. The outliers, on inspection, correspond to bad gain solutions computed from bad data with RFI in the beam, bad weather, or the Sun in the beam. Within data from one month, the standard deviation of the instrumental delays varies from $\sim\SI{1.6}{\nano\second}$ from the set of all gains computed and $\sim\SI{0.064}{\nano\second}$ from the set of gains with the bad gains discarded (over all antennas).

\section{Operations}\label{sec:ops}

The system is fairly automated. It has worked almost uninterrupted for months with the exception of a period from $\sim3$ months between July and November 2021 when the power to the site had been disrupted due to a failed transformer. Within a year of operations, we have acquired data for two high S/N FRBs. Operations involve checking if all services are working and moving daily calibrator data off-site. 

\subsection{Diagnostics and Input Flagging}

The feeds and the RFoF transmitters are not indoors and are prone to physical movement from winds as well as daily weather changes. We indirectly assess the health of the analog chain by measuring the level of the recorded analog signal by computing the root mean square (RMS) of the raw ADC snapshots. We can compute the ADC count RMS, which is very low for a disabled input and very high, close to saturating all bins in the presence of strong radio frequency interference (RFI). Additionally, we computed the spectrum over time from snapshots and inspected it for any tones due to errant analog chain oscillations. Plotting and visually inspecting a histogram of raw ADC counts is another useful diagnostic visualization wherein counts bins close to zero correspond to a disabled input, a Gaussian distribution is expected from random noise for a typical input, and a Gaussian with steep edges at the highest bin numbers is expected for saturated inputs typically from a bright sine wave in the data from an oscillating input. These checks are critical for determining inputs to include in interferometric operations and for compliance with the radio-quiet zone site regulations at the Green Bank Observatory. 

Periodically, data from every digital input are inspected for peculiar RFI or analog electronics artifacts. Some combined signal ``quicklooks" involve computing the incoherent sum, i.e. the sum of the power of and discarding predetermined "bad" inputs \seefig{fig:feb18cs_0}. For meaningfully assessing the data, RFI (typically caused by over-the-air television station broadcasts and common band radio from vehicles passing by the observatory) needs to be excised. Some of these RFI frequency bands are persistent and can be visually flagged by inspecting several days of data. We automate this by setting a threshold for the standard deviation of the noise computed using the median absolute deviation for each frequency\footnote{99.9\% of random data are contained within three standard deviations of the mean. As the signal from RFI is much stronger than the background noise, we expect it to fall outside the standard Gaussian distribution. A simple RFI detector would detect signals above three standard deviations from the mean as RFI. However, the mean and the standard deviations skew towards outliers in the data. The more RFI is present, the larger the upper and lower bounds will be, and this ends up with RFI not being identified. To mitigate this,  we can use the median absolute deviation as a statistic. It is calculated for a set of random variables $X$ as $\mathrm{MAD}=\operatorname{\mathrm{Median}}|X-\mathrm{Median}(X)|$. The MAD of normally distributed data can be used to determine the standard deviation $\sigma_{X}=1.4826 * \mathrm{MAD}$. We can then use this standard deviation $\sigma_{X}$ as a metric for the RFI  threshold to remove RFI or compute the median of the MAD in every frequency channel. Any signal above this Median(MAD) can be classified as RFI. This has been empirically demonstrated on many datasets from TONE. This method removes RFI frequency channels where RFI dominated more than 50\% of the channel and is quite aggressive.}.

\subsection{First light}

\begin{figure}[htp]
    \centering
\includegraphics[width=\textwidth,keepaspectratio]{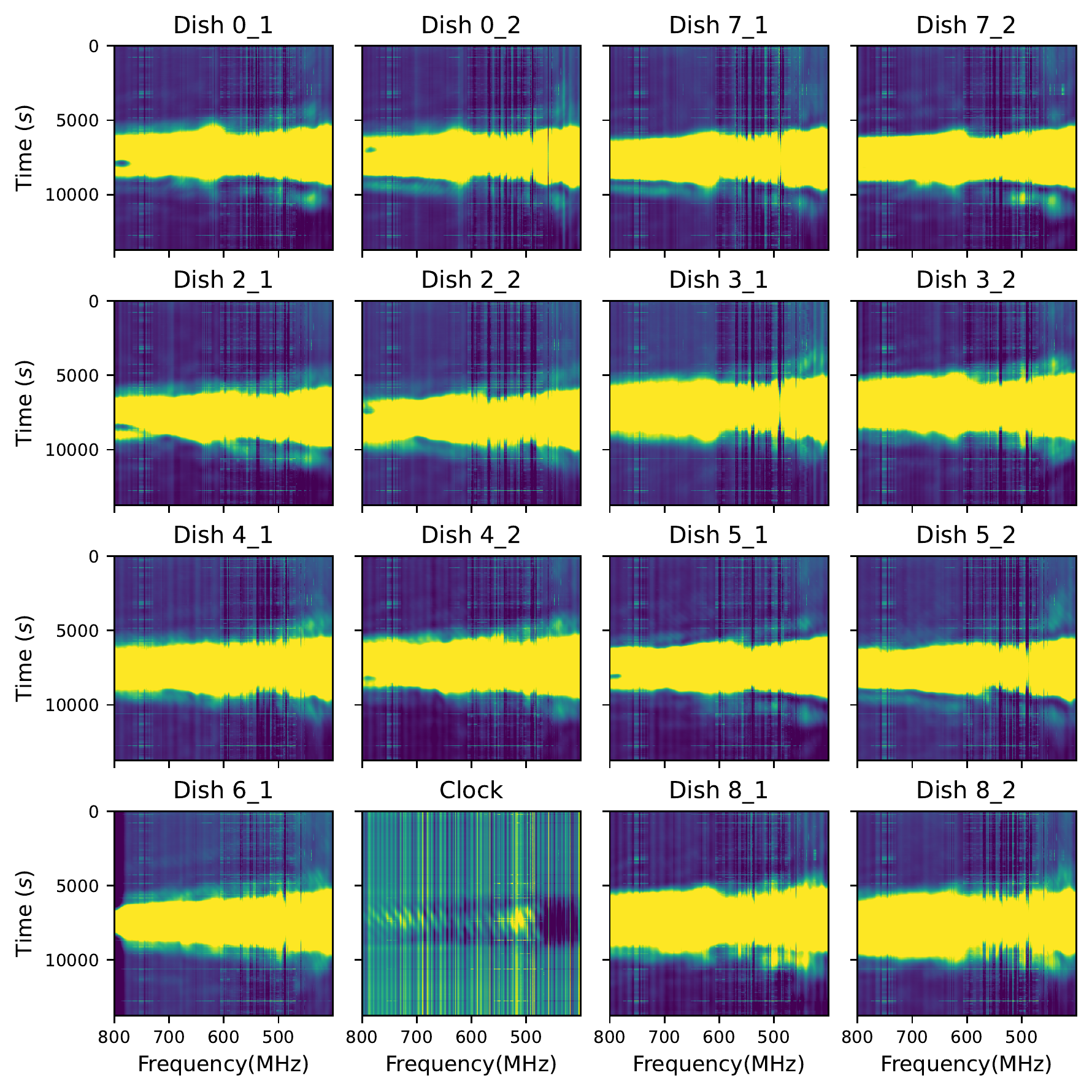}
    \caption[A drift scan of the sun]{A drift scan of the Sun. The panels show ``waterfalls'' of the intensity of the data integrated to \SI{2.7}{\second}. The bright yellow indicates when the Sun passes through the beam. The vertical stripes are consistent with RFI from television stations and the variations across frequency are from the analog chain. The panel labeled ``Clock'' shows the data from the ADC input with the clock (not a sky signal); the apparent frequency structure is due to cross-talk from the neighboring ADC inputs.}
    \label{fig:driftscansun}
\end{figure}

\begin{figure}[htp]
    \centering
    \includegraphics[width=\textwidth,keepaspectratio]{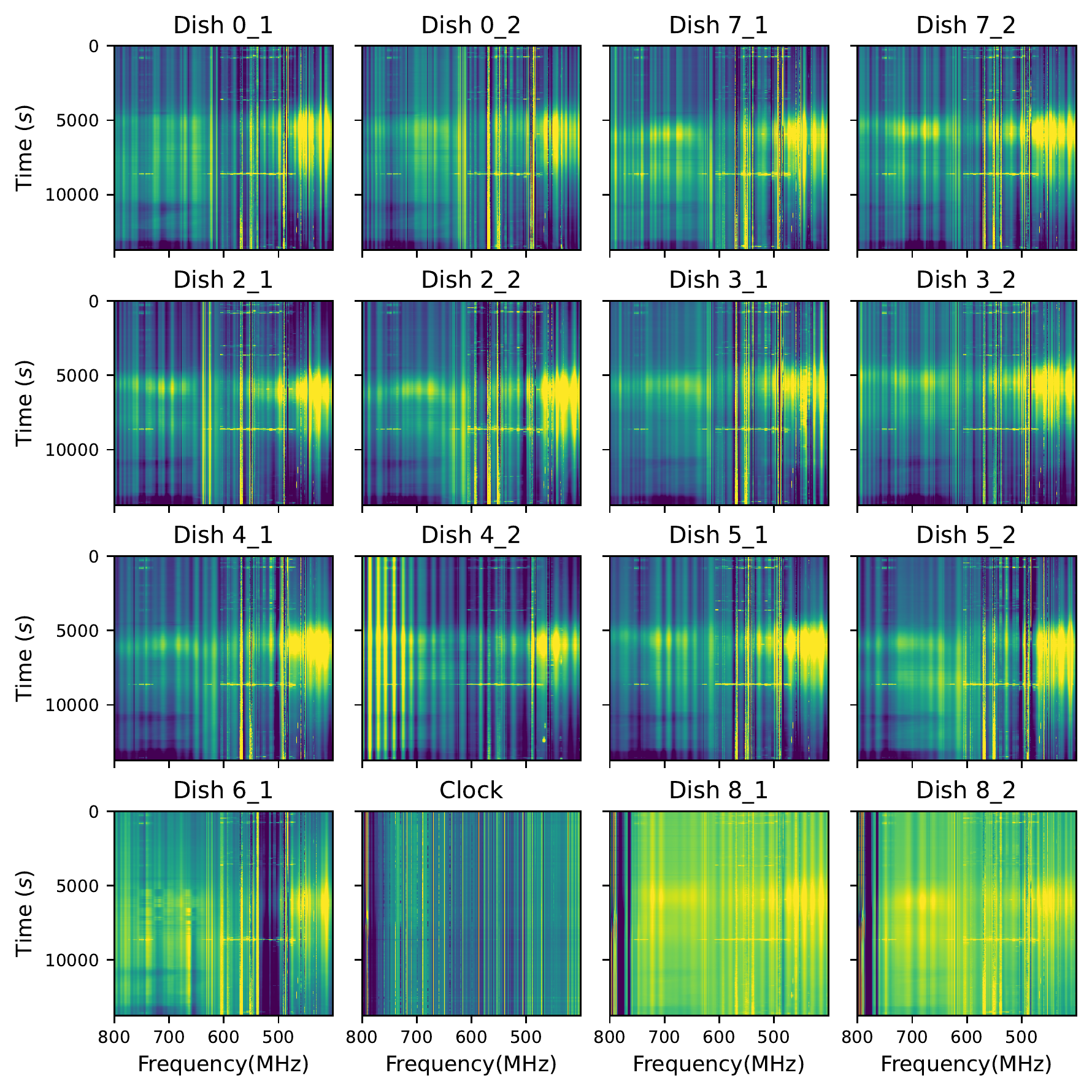}
    \caption[A drift scan of Taurus A/Crab Nebula/M1]{A  drift scan of the Crab Nebula (Taurus A). The panels show ``waterfalls'' of the intensity of the data integrated to \SI{2.7}{\second}. The bright yellow indicates when M1 passes through the beam. The vertical and some horizontal stripes are local RFI from television stations and/or maintenance equipment. The variations across frequency are from the analog chain. The panel labeled ``Clock'' shows the data from the ADC input with the clock, not the analog signal. The line at \SI{790}{\mega\hertz} is the clock signal aliased from \SI{10}{\mega\hertz}.}
    \label{fig:driftscan}
\end{figure}

The array achieved its first light in August 2020 with the real-time triggering system implemented and working by the end of October 2020. The first attempts at characterization involved running the system in the correlator mode of the ICE board and capturing the correlated visibilities from all the inputs. The total intensity data from all the inputs were used to assess the beam and the time of transit. This was used for adjusting the pointing of the dishes. Representative drift scans of the Sun and Taurus A are shown in Figure~\ref{fig:driftscansun} and Figure~\ref{fig:driftscan} respectively.

\begin{figure}[htp]
    \centering
    \includegraphics[width=0.9\textwidth,keepaspectratio]{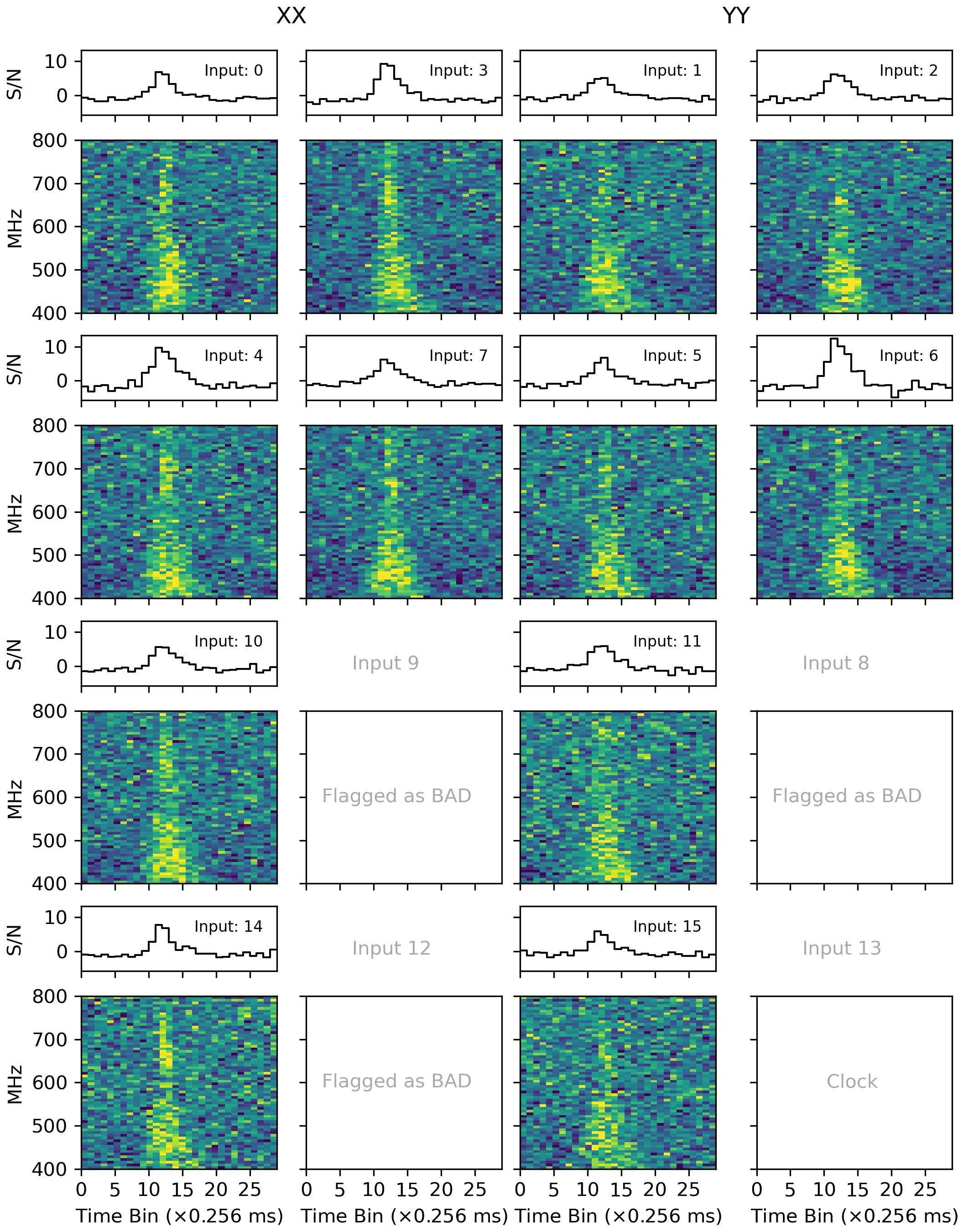}
    \caption[Crab pulsar giant pulse from February 18]{Crab pulsar giant pulse from February 18, 2021. The $4 \times 4$ grid of panels shows the power of the incoherently-dedispersed data with 16 binned frequency channels and time samples binned to \SI{0.256}{\milli\second} in the bottom subplot and a sum of all the frequency channels on the top. Inputs 8,9 and 12 do not show data because they were determined to be bad (the respective feeds malfunctioning and sending no signal at the time), while Input 13 was connected to the clock signal. The left two columns are grouped since the polarizations of the inputs are aligned and the right column is the orthogonal polarization.}
    \label{fig:feb18wfalls}
\end{figure}

TONE detected its first Crab pulsar giant pulse in the data saved after receiving a trigger from CHIME/FRB on February 10, 2021. The data from a bright pulsar from February 18 2021 are shown in Figure~\ref{fig:feb18wfalls}. The incoherent sum of these data is shown in Figure~\ref{fig:feb18cs_0}.

\subsection{Beamforming}\label{subsec:beamforming}
\begin{figure}[htp]
    \centering
    \includegraphics[width=\textwidth,keepaspectratio]{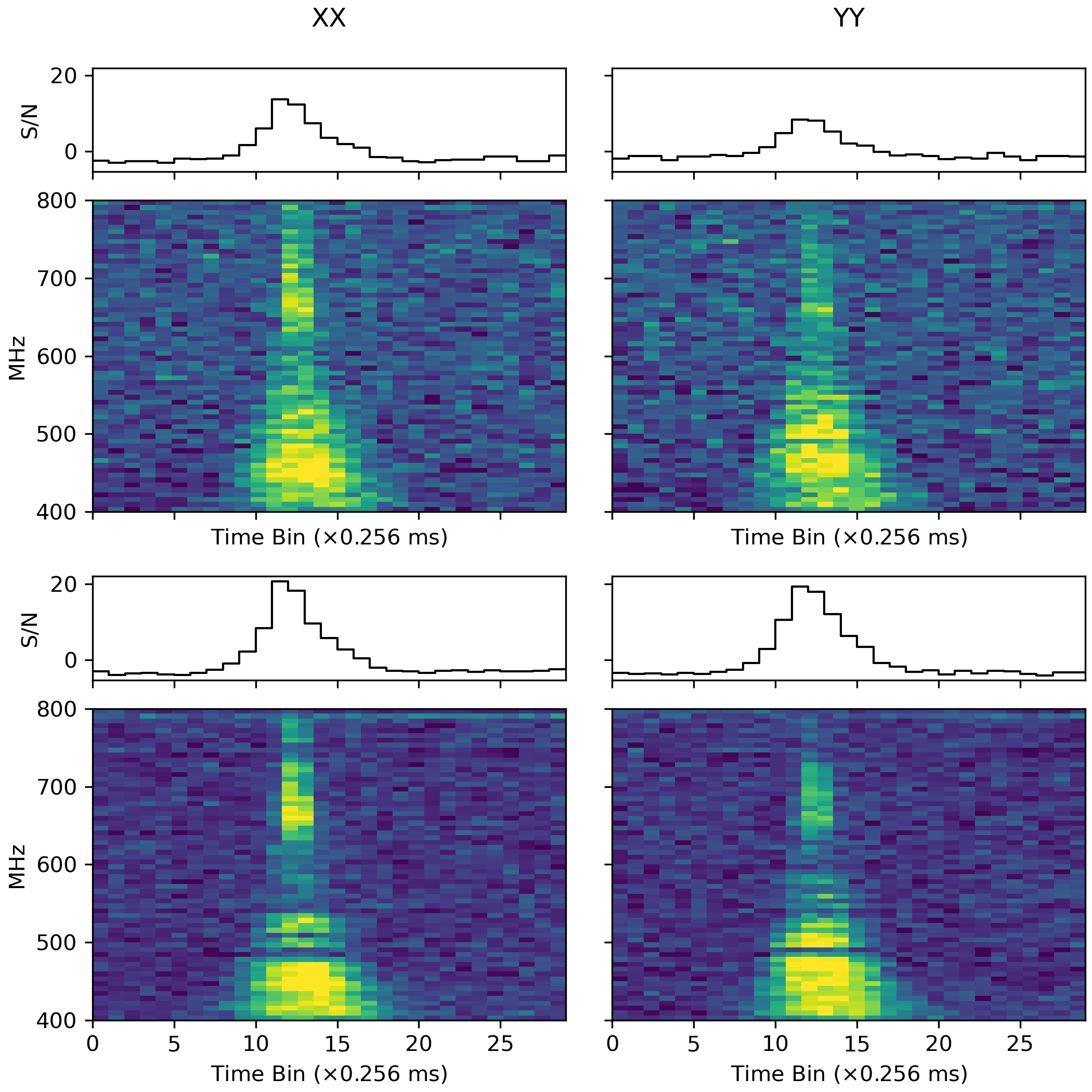}
    \caption[]{Crab pulsar giant pulse from February 18, 2021. Top row: The incoherent sum over antennas. Bottom row: The beamformed (coherently added) data. Each panel shows the power of the incoherently-dedispersed data with 16 binned frequency channels and time samples binned to \SI{0.256}{\milli\second}. The left and right columns correspond to the two orthogonal polarization groups.}
    \label{fig:feb18cs_0}
\end{figure}

To operate TONE as a phased array, we applied the gains calculated in \S\ref{sec:cal} and apply phases to steer the synthesized beam away from the zenith to the source interest in offline data analysis. Similar to the array calibration (described in \S\ref{sec:cal}), which is done separately for each polarization group, we synthesize beams using each polarization group separately. We use pipelines adapted from CHIME/FRB~\citep{chimebaseband} to form synthesized beams from the two polarization groups: we denote the tied-array data as $E_{1,2}(\nu_i,t)$, where the $1,2$ denote the two orthogonal polarization groups. Additionally, we implemented a parallelized GPU-based beamformer for independent verification and redundancy. In Figure~\ref{fig:feb18cs_0}, we compare the incoherently summed power from all antennas to the coherently summed (beamformed) power using data from a bright Crab pulsar giant pulse observed on February 18, 2021. The signal-to-noise ratio of the Crab pulsar pulse between the two waterfall plots increases by a factor of $\approx \sqrt{N}$ for $N = 6$ (the number of antennas combined). The same dataset can be used to map the point spread function of the synthesized beam, which is shown in Figure~\ref{fig:synthesized}.

\subsection{Crosstalk Characterization}
In beamformed observations towards unpolarized sources, we can quantify the level of crosstalk between the two orthogonal polarization groups of the phased TONE array. We use off-pulse observations of Crab pulsar giant pulses to form a beam pointing at Taurus A. The brightness of Taurus A means that with $\approx \SI{1}{\milli\second}$ of data, it is possible to calculate with high fidelity the complex cross-correlation coefficient between $E_{1,2}(\nu_i,t)$ to characterize cross-talk within the feed groups. In the absence of cross-talk, we would expect the complex cross-correlation coefficient $r(\nu_i)$ to be zero.

We detect significant phase structure across the band within a single transit of Taurus A, corresponding to a non-negligible level of cross-talk between the two polarization groups. We use the data from multiple transits over a two-week period to demonstrate that the phase and amplitude of the leakage are stable over time: we plot the phase and amplitude of the correlation coefficient $r(\nu_i)$ computed between the beamformed baseband data from both polarization groups, which we define in Eq.~\ref{eq:xy_corr} as

\begin{equation}
    r(\nu_i) = \dfrac{\sum_t E_1(\nu_i,t)\overline{E_2}(\nu_i,t)}{\sqrt{\sum_t|E_1(\nu_i,t)|^2}
     \sqrt{\sum_t|E_2(\nu_i,t)|^2}}.
     \label{eq:xy_corr}
\end{equation} 

\begin{figure}[htp]
    \centering
    \includegraphics[width=\textwidth]{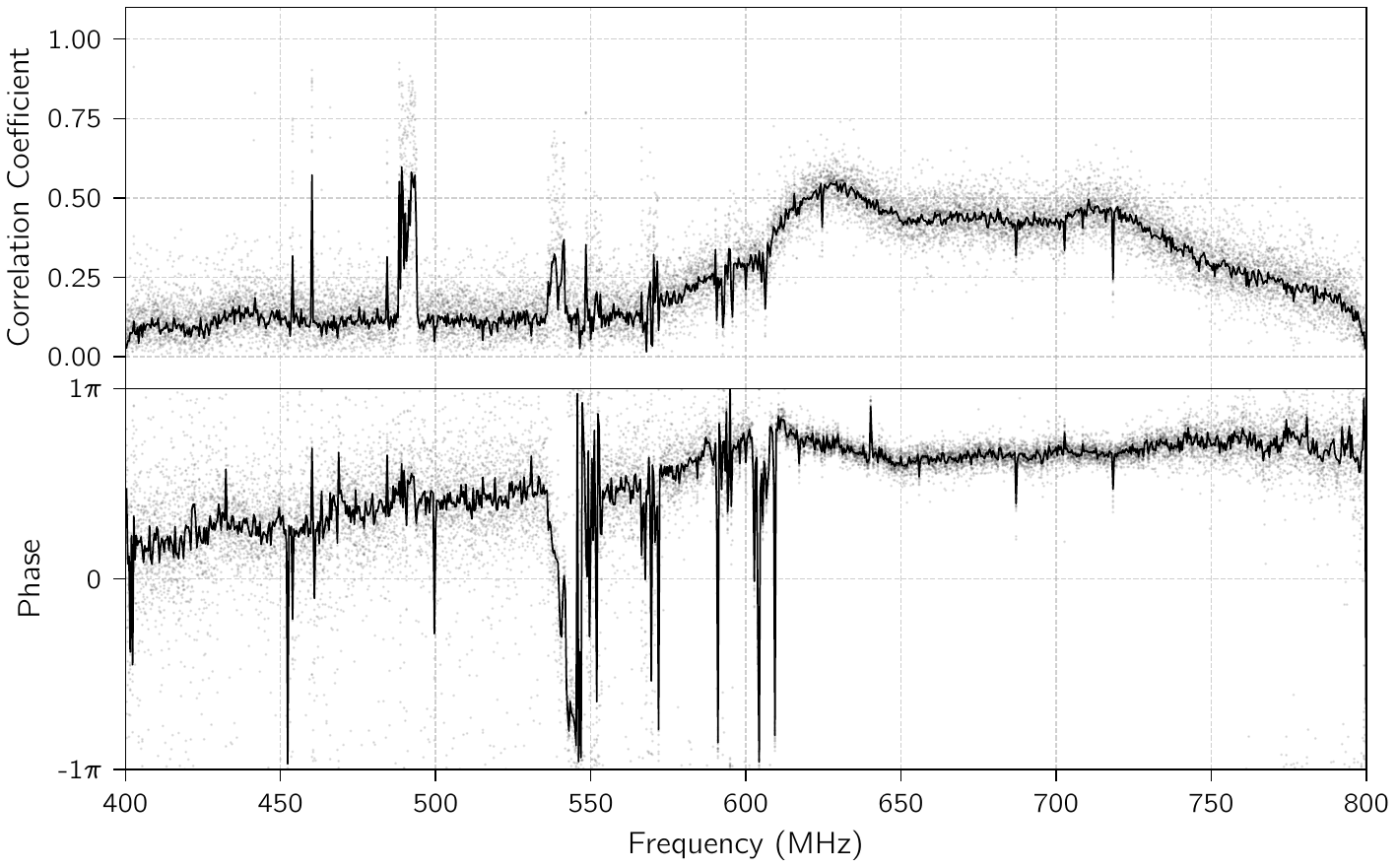}
    \caption{The amplitude and phase of the cross-correlation coefficient between the two polarization groups as a function of frequency, for individual transits of a bright, unpolarized source (Taurus A) and for the stacked transits (solid). In the upper half of the band, $\approx 50\%$ of the power in XX and YY is highly correlated due to cross-talk between the two polarization groups. The impedance mismatch at $\sim 550-\SI{650}{\mega\hertz}$ is evident not only in the SEFD (see Figure~\ref{fig:test}) but also as a highly unstable relative phase between polarization groups.}
    \label{fig:xy_phase}
\end{figure}

\begin{figure}[htp]
\centering
  \includegraphics[width=\textwidth]{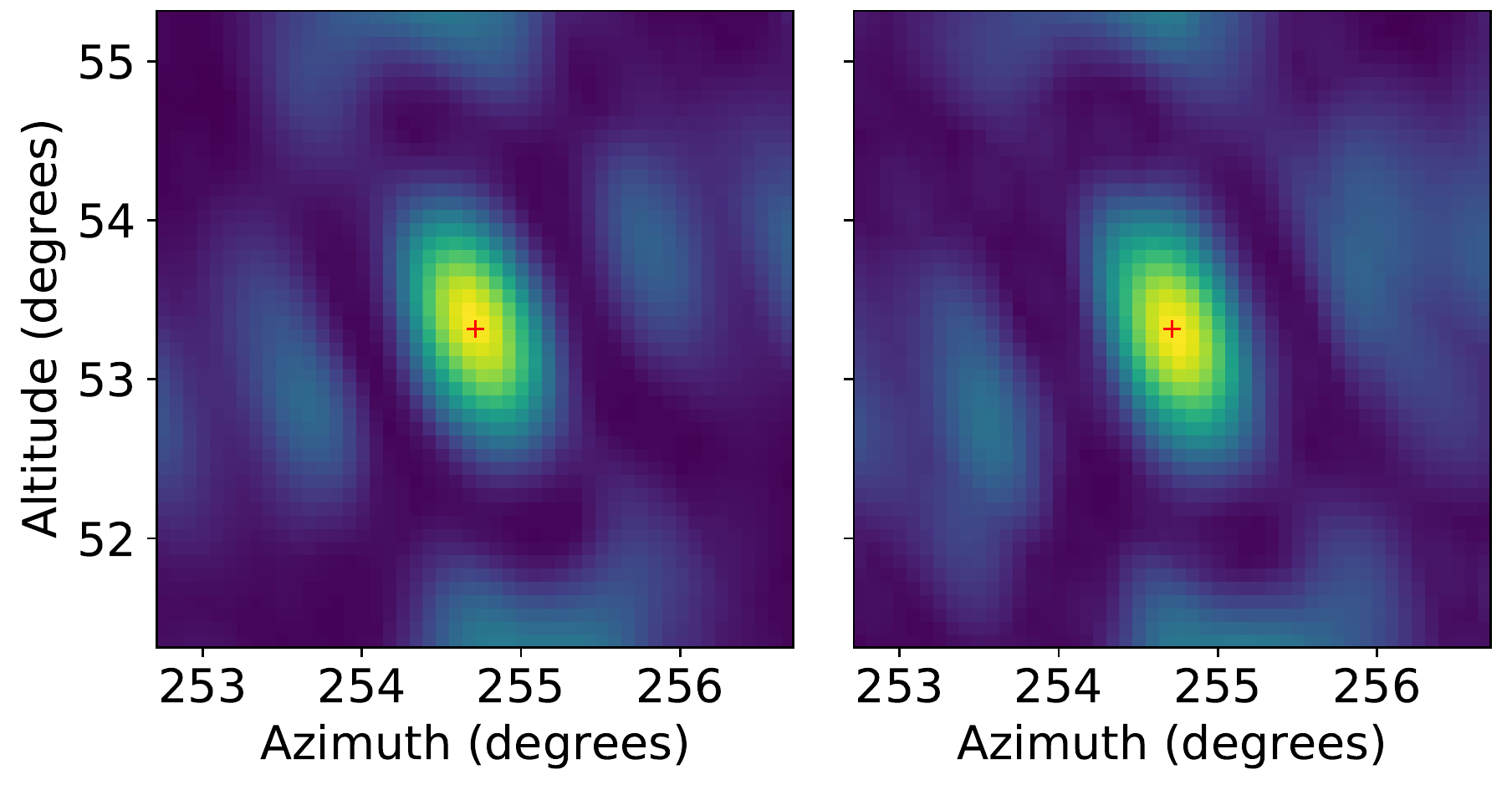}
  \caption[Grid of S/Ns of the Crab pulsar on the sky]{A map of the synthesized beam of TONE. We plot, as a function of pointing center $\pm \SI{2}{\deg}$ around the Crab pulsar's true position (red plus), the S/N of the Crab pulsar giant pulse signal summed over all frequencies from the February 18 giant pulse (see Figure~\ref{fig:feb18cs_0}). The two plots correspond to the beams synthesized from two orthogonal polarization groups.}
  \label{fig:synthesized}
\end{figure}

We show the data from individual days (faint black points) and from all the transits stacked or co-added (solid, black points) in Figure~\ref{fig:xy_phase}.
From the slope of the phase across the band ($\approx 1$ wrap across the band), we can see that signals polarized along the direction of group 1 leak into the group 2 signal path with a delay on the order of a nanosecond and that the cross-polarization leakage is on the order of $50\%$ at higher frequencies. This leakage can complicate measurements of the ionospheric phase residuals, which can affect localizations at the sub-arcsecond level.

\section{Triggered VLBI Observations with CHIME/FRB} \label{subsect:xcorr}
We correlate beamformed baseband data from Crab giant pulses, as demonstrated initially in~\cite{2022AJ....163...48M}. Because the relative orientation of the feeds in TONE relative to those at the CHIME telescope is $\approx 30^\circ$, we form circular polarizations out of the data from the linear feeds at both stations. The relative phase between the polarization groups at both stations varies spatially over the primary beam in both telescopes, but this corresponds to a delay of $\lesssim 1$ ns and is stable over time (see Fig.~\ref{fig:xy_phase}). This is subdominant to current clocking errors, which are the largest systematic errors for FRB localization. Therefore, at this time, we do not attempt to perform polarization calibration to TONE autocorrelations or CHIME-TONE cross-correlations.

The VLBI correlation is performed as follows. For each frequency channel of each beamformed baseband data centered at frequency $\nu_i$, we fringestop a short scan of data, compensating for the known delay towards the proper-motion corrected position of the Crab pulsar as measured in~\citet{lobanov2011vlbi} at the epoch of our observations:

\begin{equation}
    \alpha_\mathrm{J2000} = 83.6330379, \qquad \delta_\mathrm{J2000} = 22.0145018.
\end{equation}

Here we report both the J2000-epoch right ascension ($\alpha$ and $\delta$) in decimal degrees, omitting error bars because our systematic errors dominate the astrometric uncertainty from extrapolating the proper motion to the present epoch, as well as the measurement uncertainty in that work. The start time of the data differs from channel to channel because of dispersion delay, and the scans are centered on the pulse's fiducial TOA and fiducial DM at the central frequency of each channel. This frequency-dependent gating in each channel essentially applies incoherent dedispersion to each frequency channel and reduces background noise.

It is possible to further concentrate the signal along the time axis since the pulse width is shorter than the intra-channel smearing timescale. We apply coherent dedispersion to data from both stations via the transfer function $\exp(2\pi i H(\Delta \nu))$, where

\begin{equation}
    H(\Delta \nu) = \dfrac{K_\mathrm{DM} \mathrm{DM} \Delta \nu^2}{(\nu_i + \Delta \nu)\nu_i^2}.
\end{equation}

\begin{figure}[htp]
    \centering
    \includegraphics[width =\linewidth]{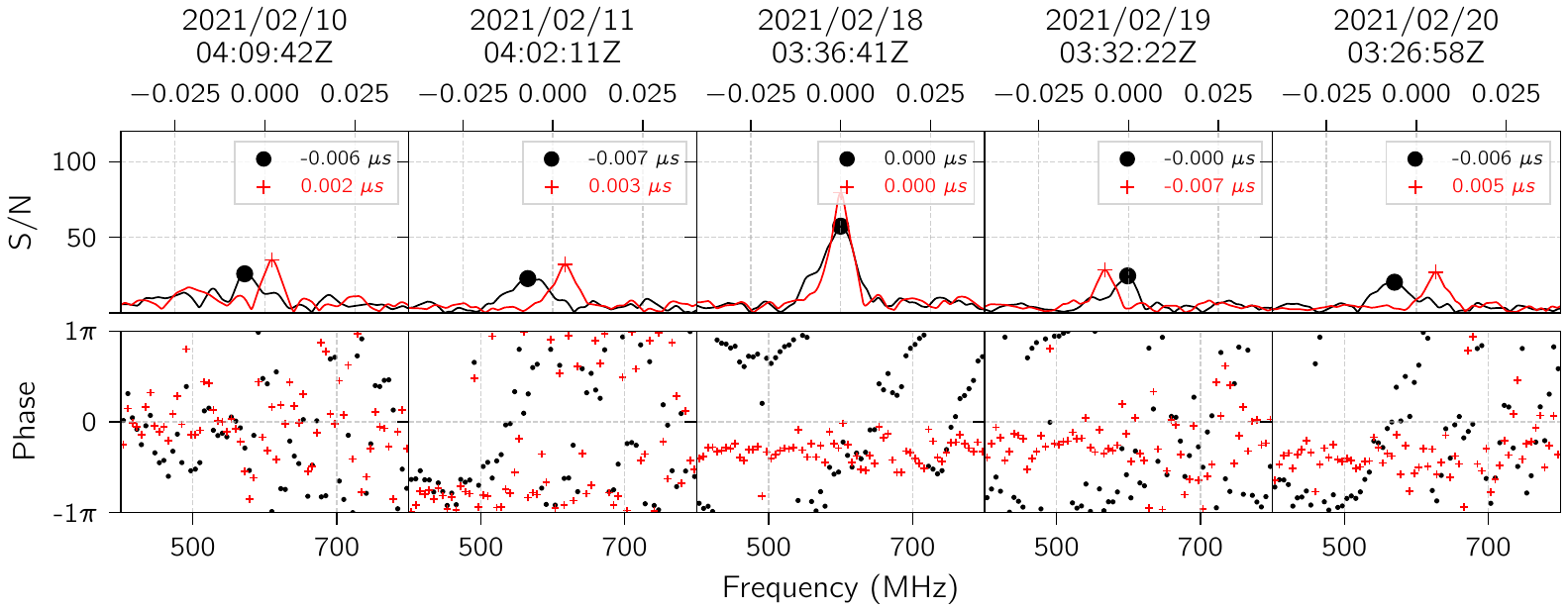}
    \includegraphics[width =\linewidth]{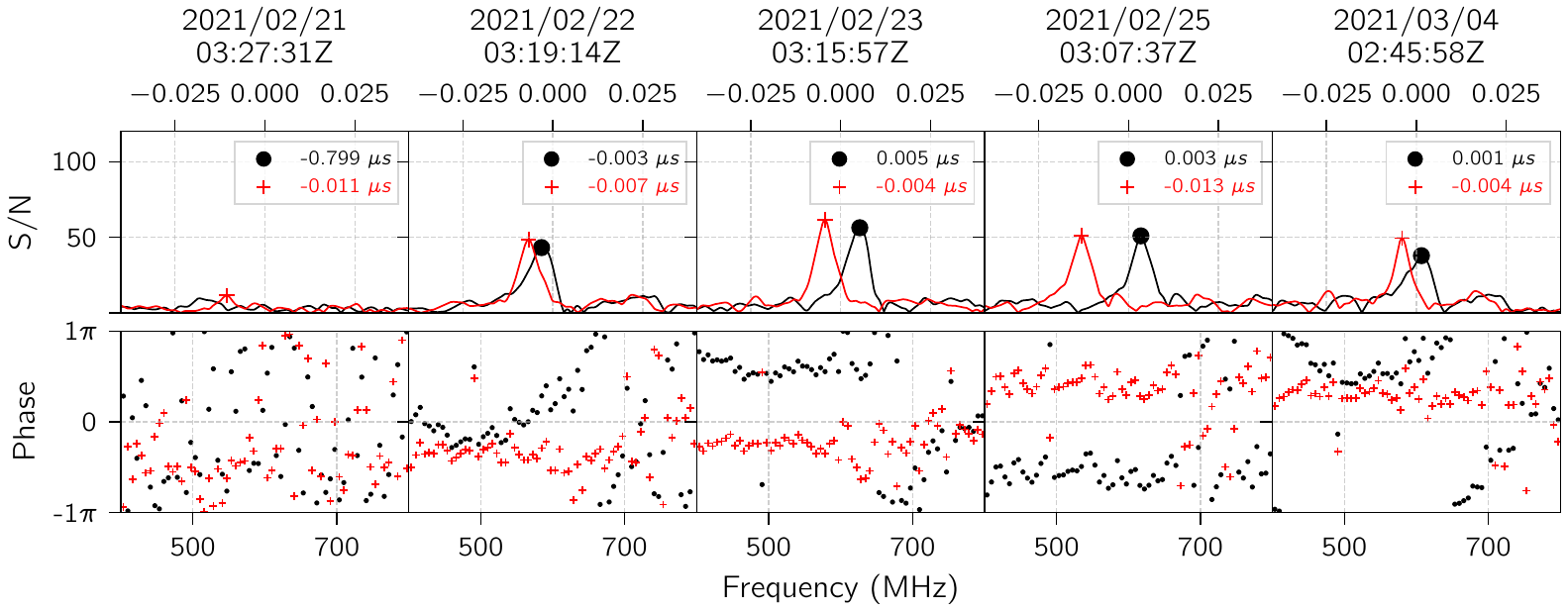}
    \includegraphics[width =0.6\linewidth]{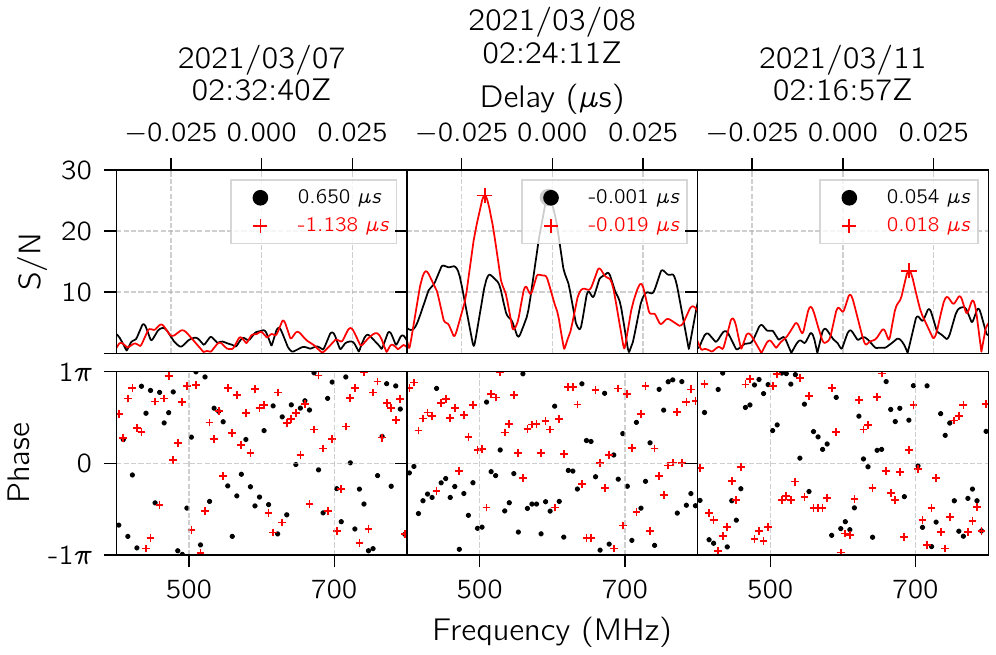}
    \caption{Cross-correlation visibilities for a sequence of Crab pulsar giant pulses, collected in February and March 2022. Rows labeled ``S/N'' show the cross-correlation signal-to-noise ratio as quantified by Eq.~\ref{eq:rho}, i.e. the absolute magnitude of the Fourier transform of the respective visibility phases, which are shown in the rows labeled ``Phase''. Black and red traces show before and after removing the best-fit total electron content (TEC) correction, whose $\varphi \propto 1/\nu$ frequency dependence can be directly measured over our $400-\SI{800}{\mega\hertz}$ band (see data from e.g. 2021/02/20) The cross-correlation S/N as a whole decreases over time (see bottom row), between successive manual repointings of the TONE dishes.}
    \label{fig:fringes}
\end{figure}

\noindent Here, $\SI{-195.3125}{\kilo\hertz} \leq \Delta \nu < \SI{195.3125}{\kilo\hertz}$ refers to the difference between the true frequency and $\nu_i$, $K_\mathrm{DM}=\SI{1/2.41e-6}{\second\mega\hertz\squared\per\parsec\centi\meter\cubed}$ 
\cite{2004hpa..book.....L,2020arXiv200702886K}, and the DM is the fiducial DM used to line up the pulse within its frequency-dependent gate. It takes on values between \SIrange{56.746}{56.76}{\parsec\per\centi\m\cubed} for all of the pulses observed. The width of the gate is between $\SIrange{200}{400}{\micro\second}$, chosen by inspecting each giant pulse's morphology. We also correlate 12 off-pulse gates to estimate the variance $\sigma_i^2$ as a function of the frequency channel. The cross-correlation visibilities resulting from this procedure are shown in Figure~\ref{fig:fringes}. 

\subsection{Localization}\label{subsect:localization}
As a first step towards localizing single pulses, we measure group delays over the CHIME-TONE baseline using the phases in Sec.~\ref{subsect:xcorr}. As an initial search for fringes, we divide the visibilities $V_i$ by $\sigma_i$ to perform signal-to-noise normalization, and Fourier transform the visibilities over the frequency axis. This corresponds to searching for delays by evaluating the cross-correlation function $\rho(\tau)$ over a range of $[-1.28,+1.28)$ \si{\micro\s}. We visually inspect the phases by unwrapping them by the best-fit delay. The result wraps slowly over the frequency band and can be binned by a factor of 16 for visualization.

If fringes are successfully found this way, the data are further analyzed. Over a month of operations, we collected 13 sets of Crab pulsar giant pulse fringes this way. To compensate for the ionosphere, we perform a grid search over differential slant total electron content (TEC) values between $\Delta\mathrm{DM} \in [-10,+10)$ TEC units, corresponding to a dispersion measure difference of $\pm 3.2\times10^{-6}$ pc cm$^{-3}$ between stations. The TEC values we measure all fall into this range. We optimize
\begin{equation} 
\rho(\tau,\Delta\mathrm{DM}) = \left| \sum_{i} V_i \exp(-2\pi i (\nu_i \tau + K_\mathrm{DM} \Delta\mathrm{DM} / \nu) )/ \sigma_i \right|
\label{eq:rho}
\end{equation}
over all values of $\tau,\Delta\mathrm{DM}$. 

To calibrate our data, we first apply a delay correction of $\tau = \SI{0.473}{\micro\second}$, measured from 2021/02/18 data -- our brightest Crab pulsar pulse -- as our reference point. The stability and magnitude of this large offset over several months lead us to interpret the origin of this delay as the difference in the analog cable lengths between TONE and CHIME. We are unaware of other effects in our instruments which can induce delays of hundreds of nanoseconds. Furthermore, the size of this effect is consistent with the static delay expected from the instrumental cable lengths. This static delay correction does not affect our planned analysis, which is only sensitive to fluctuations in this delay over 24-hour timescales.

We next apply a phase shift to $V_i$ corresponding to the value of $\Delta \mathrm{DM}$ that maximizes Eq.~\ref{eq:rho}: $\varphi^{\mathrm{TEC}}_i = 2\pi K_\mathrm{DM} \Delta \mathrm{DM} / \nu_i$.  This removes dispersive (i.e. instrumental and ionospheric) delays; any remaining shifts are non-dispersive. The effect of performing this correction on the phases can be seen in the black (pre-correction) and red (post-correction) phase curves shown in Figure~\ref{fig:fringes}. In the upper panel (labeled S/N), we have plotted $\rho$ as a function of $\tau$ over a range of $\SIrange{-0.04}{+0.04}{\micro\second}$. After correcting for the ionospheric phase, it can be seen that the peak is, generally speaking, taller and more symmetric; this is most evident in the data from 2021/02/18 (our brightest Crab pulsar pulse). In the lower panel (labeled ``Phase''), the reason for this improvement can be seen in visibility space: the curvature in the visibilities as a function of frequency has been measured and removed (e.g., data from 2021/02/18 and 2021/02/22). After unwrapping a group delay, the remaining phase residuals are more constant as a function of frequency, resulting in the observed boost. Applying these best-fit TEC corrections shifts the geometric delay by $\sim \SI{10}{\nano\second}$. 

\section{Empirical determination of localization error}\label{sect:systematicerror}
\begin{figure}[htp]
    \centering
    \includegraphics[width=0.9\textwidth]{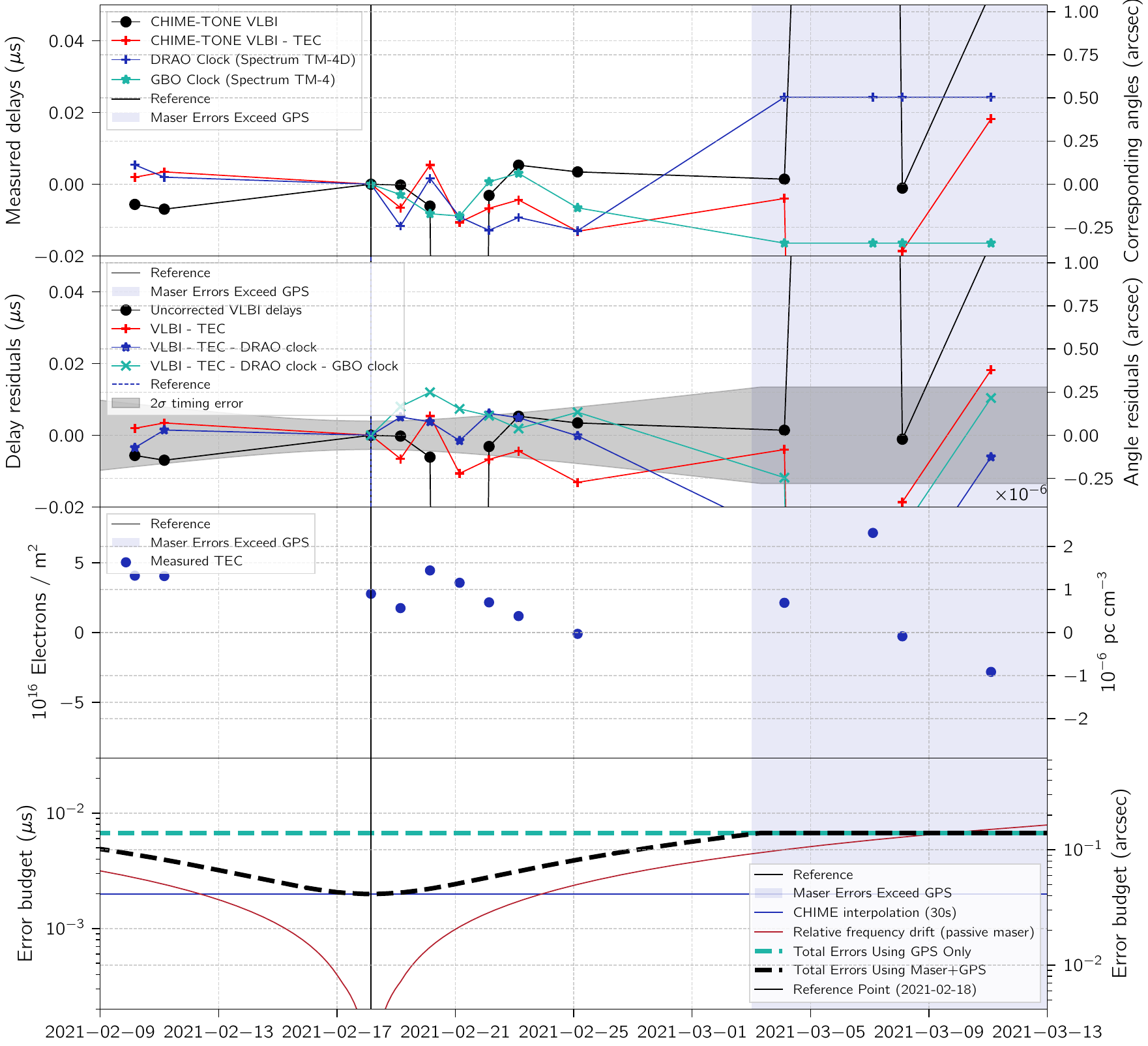}
    \caption{Empirical measurements of delay residuals as a function of time for an observing run in February-March 2021. \textbf{Top:} We measure VLBI delays in single-pulse Crab pulsar data. compensating only for geometric delays (black circles) and compensating for geometric delays and the best-fit slant TEC in each observation (red stars). Local clock corrections are computed at DRAO (blue plusses) and at GBO (green crosses). It can be seen visually that the local clock corrections roughly trace the measured delays. \textbf{Second:} We apply clock corrections to measure residual delays. Uncorrected delay residuals (black and red, same as the top plot) are shown for comparison with the corrected delay residuals after applying DRAO clock corrections (blue) and both DRAO and GBO (green) clock corrections. \textbf{Third:} The measured differential Total electron Content (TEC) between the line of sights of CHIME and TONE. \textbf{Bottom:} For comparison with theory, we compare the residuals to expected errors from the short-timescale jitter of the GPS clock at CHIME and the long-timescale relative drift of the masers at both stations.}
    \label{fig:delay_residuals_feb}
\end{figure}

To validate these multi-day delay measurements from the Crab pulsar, we plot both the uncorrected and corrected delays in the top panel of Figure~\ref{fig:delay_residuals_feb} as a function of time in black and red respectively. To monitor changes over time, we reference these two time series to zero at the TOA of the Crab pulsar pulse measured on 18 February 2021, since it is the brightest pulse in our set, and monitor deviations from this reference to characterize time-varying systematic shifts.

The largest time-varying shifts after compensating for the ionosphere are the local clock corrections from the TONE and CHIME. As mentioned in Sec.~\ref{subsec:ice_board}, TONE's primary clock is a Spectrum TM--4 unit whose root mean squared (RMS) deviation is $\approx \SI{6}{\nano\second}$. The Spectrum TM--4D unit at CHIME, however, has slightly worse performance, with an RMS deviation of $\approx \SI{20}{\nano\second}$ which we attribute to an additional distribution amplifier in the signal chain between the GPS receiver and the CHIME F-engine (not to be confused with the distribution amplifier which distributes the maser itself in Figure~2 of~\citet{2022AJ....163...48M}). To characterize clock-related contributions to the delay residuals, we measure and plot the clock corrections at both the CHIME and TONE using the pipeline described by~\citet{2022AJ....163...48M}. If the only two systematics present were the clocks and ionosphere, we would expect that the VLBI delays, after correcting for the ionospheric fluctuations (``VLBI - TEC'' time series, shown in red in the top panel of Figure~\ref{fig:delay_residuals_feb}), exactly trace the sum of the DRAO and GBO clock corrections, since only geometric delays are applied in the VLBI correlator.

We apply the clock corrections from the VLBI-TEC time series (red stars in the top panel of Figure~\ref{fig:delay_residuals_feb}) and find partial success. Subtracting the larger contribution (the DRAO clock, blue crosses in the top panel of Figure~\ref{fig:delay_residuals_feb}) from the VLBI-TEC time series gives residuals that are significantly reduced from the delay residuals before applying this correction (blue crosses in the middle panel of Figure~\ref{fig:delay_residuals_feb}). However, when we further apply the TONE clock corrections (green X's in the top panel of Figure~\ref{fig:delay_residuals_feb}), the residuals (green X's in the middle panel of Figure~\ref{fig:delay_residuals_feb}) grow again.

The agreement of the blue, but not the green, delay residuals shows that we have successfully applied the local clock correction at DRAO but not the local clock correction at TONE. We note that even without applying the TONE clock correction (i.e. using the Spectrum TM--4), our delay residuals are on the order of $\approx \SI{10}{\nano\second}$, which corresponds to an $\approx \SI{200}{\mas}$ single-pulse localization. Our best explanation for this is that some additional systematics are present in the TONE clock distribution pipeline which is not yet understood.

We find these empirical residuals almost consistent with the theoretical expectation for the errors in our clocking system as a function of time (bottom panel of Figure~\ref{fig:delay_residuals_feb}). To make this comparison, we model two contributions to the delay residuals from the clock system: the jitter induced over the 30 seconds between successive readouts of the DRAO maser at CHIME, as well as the long-term relative frequency drift between the two masers. These contributions are shown as blue and red traces in the bottom panel of Figure~\ref{fig:delay_residuals_feb}. The short-term jitter is characterized by the Allan deviation of the Spectrum TM--4D at CHIME as measured on 30-second timescales. The relative frequency drift between the two masers starts at zero, and increases as more time elapses since the 18 February calibration. We estimate the size of the relative frequency drift of the two masers by assuming that their Allan variances are similar. We then add the Allan variance of a single maser in quadrature with itself (i.e. multiplying the Allan deviation by $\sqrt{2}$). The Allan deviation measurements we use are shown in Figure~1 of~\citet{2022AJ....163...48M}; note that this does not include any effects of the signal chain used to distribute the GBO maser signal from the GBO maser itself to TONE. That signal chain includes an active RFoF transmitter and receiver system, and may plausibly induce additional variations in the delay residuals on shorter timescales than the intrinsic fluctuations in the actual maser.

We add the short-term jitter and the long-term drift contributions in quadrature to estimate our 1$\sigma$ total error budget (black dotted line in the bottom panel of Figure~\ref{fig:delay_residuals_feb}). On short timescales ($\lesssim \SI{30}{\second}$), we are dominated by DRAO maser readout at CHIME, while on longer timescales, the relative drift between the oscillation frequencies of the two masers dominates ($\lesssim \SI{24}{\hour}$). On the longest timescales ($\approx$ 2 weeks), we enter the regime where the fact that the masers are drifting independently of each other becomes significant (red line in the bottom panel of Figure~\ref{fig:delay_residuals_feb}). 

In this regime (blue shaded area of Figure~\ref{fig:delay_residuals_feb}), it is more accurate to use GPS clocks, since the clocks at both stations are actively synchronized by GPS satellites. The absolute precision of GPS clocks is characterized simply by an RMS deviation from true time. For TONE, the RMS deviation is $\SI{3.4}{\nano\second}$ (Figure~\ref{fig:adev}). For CHIME, the RMS deviation is \SI{6}{\nano\second} (see Figure~1 of~\cite{2022AJ....163...48M}). The delay residuals from using GPS clocking are therefore somewhat bracketed at $1\sigma$ by the RMS of the GPS clocks added in quadrature ($\approx \SI{7}{\nano\second}$).

For comparison with the data, we shade a $2\sigma$ error budget centered at zero in the center panel of Figure~\ref{fig:delay_residuals_feb}. We use \texttt{difxcalc11} to compute $d\tau/d\alpha$ (where $\alpha$ is the right ascension) for a typical observing geometry to translate our empirical delay residuals into empirical systematic astrometric offsets. We conclude that on the TONE baseline, our best astrometric errors from empirical data are on the order of $\SI{0.04}{\arcsec}$ on short timescales, which grow to $\SI{0.2}{\arcsec}$ on long timescales due to the slow cadence of VLBI calibration.

\section{Discussion and Conclusion}\label{sect:dnc}

TONE, along with the other outrigger pathfinders ARO10 and the CHIME pathfinder, are key precursors to the full CHIME/FRB Outriggers. The three stations will be near Princeton, British Columbia, Canada, GBO, Green Bank, West Virginia, and Hat Creek Radio Observatory, Hat Creek, California giving us $\sim67$ km, $\sim3333$ km, and $\sim955$ km baselines respectively. The outrigger station at Princeton is a single \SI{20}{\meter} wide and \SI{40}{\meter} long cylindrical reflector with 64 CHIME-like cloverleaf feeds and the outriggers at GBO and HCRO will consist of single \SI{20}{\meter} wide and \SI{64}{\meter} long cylindrical reflectors equipped with 128 CHIME like cloverleaf antennas each. These sites are at various stages of construction and commissioning at the time of writing. Our work on these pathfinders has allowed us to determine and identify where we can improve our systematic errors for VLBI. 

The systematic localization error on the TONE baseline on 24-hour timescales is dominated by the TONE clock and its distribution system. Our estimate is grounded in empirical data, but it is possible that other systematics will become apparent when doing VLBI on other sources besides Crab pulsar single pulses. With the current setup, we are not able to explore the spatial dependence of these systematic errors, since observations over the CHIME-TONE baseline are essentially limited to single-pulse VLBI. While groundbreaking advances have been made by Low-Frequency Array (LOFAR) in low-frequency VLBI, the enormous data rate of CHIME limits our maximum integration duration, which makes it difficult to observe the faint VLBI calibrators identified by LOFAR~\citep{moldon2015lofar,morabito2022subarcsecond,jackson2022subarcsecond}. The small collecting area and fixed pointing of TONE at a declination of $21^\circ$ further restrict our VLBI observations at present to extremely bright, single-pulse observations, for which the Crab pulsar is essentially the only source visible. We, therefore, reserve a treatment of spatially-dependent astrometric systematics in widefield VLBI for future work. We have begun to explore these on short baselines using the CHIME Pathfinder~\cite{pathfinderoutrigger}, but the commissioning of CHIME/FRB Outriggers, whose large collecting area and cylindrical telescope geometry will make a methodical study of these systematics more possible. With that instrument, short integrations of bright point sources and single pulses from bright pulsars can be used to explore the spatial dependence of these systematics, providing a path towards localizing a large sample of FRBs at the time of detection with sub-arcsecond precision over the widest fields of view. This would allow us to greatly improve upon the current systematic error. Additionally, the Outriggers will have tracking beam capabilities. The longest period of time without a pulsar previously identified as a phase calibrator in the CHIME beam is \SI{\sim1}{\hour}; this will allow for more frequent calibrations than the day-long cadence we have explored in this work.

The TONE and ARO10 pathfinders have since successfully detected FRB 20210603A in cross correlation~\cite{2022AJ....163...65C} and we have been able to localize the source to its host galaxy. This work is detailed in \citet*{frb20210603A} and is a milestone for CHIME/FRB Outriggers, which are poised to localize hundreds of FRBs per year once they are fully commissioned.


\section*{Acknowledgments}
P.S. would like to sincerely thank José Miguel Jáuregui García, Jacob Hanni, John Makous, Thaddeus Herman, Howard Chun, John Clarke, Michael Stover and Joseph Kania for assisting with the assembly and commissioning of TONE. We are grateful for the helpful comments from Ziggy Pleunis. 

TONE is located at the Green Bank Observatory, the Green Bank Observatory facility, which is supported by the National Science Foundation, and is operated by Associated Universities, Inc. under a cooperative agreement. We would like to thank the staff at Green Bank Observatory for logistical support during the construction and operations of TONE. FRB research at WVU is supported by an NSF grant (2006548, 2018490). 

We acknowledge that CHIME is located on the traditional, ancestral, and unceded territory of the Syilx/Okanagan people. We are grateful to the staff of the Dominion Radio Astrophysical Observatory, which is operated by the National Research Council of Canada.  CHIME is funded by a grant from the Canada Foundation for Innovation (CFI) 2012 Leading Edge Fund (Project 31170) and by contributions from the provinces of British Columbia, Qu\'{e}bec and Ontario. The CHIME/FRB Project is funded by a grant from the CFI 2015 Innovation Fund (Project 33213) and by contributions from the provinces of British Columbia and Qu\'{e}bec, and by the Dunlap Institute for Astronomy and Astrophysics at the University of Toronto. Additional support was provided by the Canadian Institute for Advanced Research (CIFAR), McGill University and the McGill Space Institute thanks to the Trottier Family Foundation, and the University of British Columbia. 

C.L. was supported by the U.S. Department of Defense (DoD) through the National Defense Science \& Engineering Graduate Fellowship (NDSEG) Program. K.W.M. holds the Adam J. Burgasser Chair in Astrophysics and is supported by an NSF Grant (2008031). J.B.P. is support by the NSF MRI grant (2018490). U.-L.P. receives support from Ontario Research Fund—research Excellence Program (ORF-RE), Natural Sciences and Engineering Research Council of Canada (NSERC) [funding reference number RGPIN-2019-067, CRD 523638-18, 555585-20], Canadian Institute for Advanced Research (CIFAR), Canadian Foundation for Innovation (CFI), Thoth Technology Inc, Alexander von Humboldt Foundation, and the Ministry of Science and Technology(MOST) of Taiwan(110-2112-M-001-071-MY3). Computations were performed on the SOSCIP Consortium’s [Blue Gene/Q, Cloud Data Analytics, Agile and/or Large Memory System] computing platform(s). SOSCIP is funded by the Federal Economic Development Agency of Southern Ontario, the Province of Ontario, IBM Canada Ltd., Ontario Centres of Excellence, Mitacs and 15 Ontario academic member institutions.
%

%



\bibliographystyle{ws-jai}

\bibliography{main}
\end{document}